\documentclass[aps,prb,reprint,groupedaddress,superscriptaddress]{revtex4-2}
\usepackage{graphicx,color,bm,amsmath,amssymb,url,natbib}

\begin{document}

\title{Skyrmion Lattice Manipulation with Electric Currents and Thermal Gradients in MnSi}

\author{N.~Chalus}
\author{A.~W.~D.~Leishman}
\affiliation{Department of Physics and Astronomy, University of Notre Dame, Notre Dame, Indiana 46556, USA}

\author{R.~M.~Menezes}
\affiliation{Department of Physics \& NANOlab Center of Excellence, University of Antwerp, Groenenborgerlaan 171, B-2020 Antwerp, Belgium}
\affiliation{Departamento de F\'{i}sica, Universidade Federal de Pernambuco, Cidade Universit\'{a}ria, 50670-901 Recife-PE, Brazil}

\author{G.~Longbons}
\affiliation{Department of Physics and Astronomy, University of Notre Dame, Notre Dame, Indiana 46556, USA}

\author{U.~Welp}
\author{W.-K.~Kwok}
\affiliation{Materials Science Division, Argonne National Laboratory, Argonne, Illinois}

\author{J.~S.~White}
\affiliation{PSI Center for Neutron and Muon Sciences, 5232 Villigen PSI, Switzerland}

\author{M.~Bartkowiak}
\affiliation{PSI Center for Neutron and Muon Sciences, 5232 Villigen PSI, Switzerland}

\author{R.~Cubitt}
\affiliation{Institut Laue-Langevin, Grenoble, France}

\author{Y.~Liu} 
\author{E.~D.~Bauer}
\affiliation{Los Alamos National Laboratory, Los Alamos, New Mexico 87545, USA}

\author{M.~Janoschek}
\affiliation{PSI Center for Neutron and Muon Sciences, 5232 Villigen PSI, Switzerland}
\affiliation{Physik-Institut, Universit\"at Z\"urich, Winterthurerstrasse 190, CH-8057 Zurich, Switzerland}

\author{M.~V.~Milo\v{s}evi\'c}
\affiliation{Department of Physics \& NANOlab Center of Excellence, University of Antwerp, Groenenborgerlaan 171, B-2020 Antwerp, Belgium}
\affiliation{Instituto de F\'{i}sica, Universidade Federal de Mato Grosso, Cuiab\'{a}, Mato Grosso 78060-900, Brazil}

\author{M.~R.~Eskildsen}
\affiliation{Department of Physics and Astronomy, University of Notre Dame, Notre Dame, Indiana 46556, USA}

\begin{abstract}
The skyrmion lattice (SkL) in MnSi was studied using small-angle neutron scattering and under the influence of a radial electric current in a Corbino geometry. In response to the applied current, the SkL undergoes an angular reorientation with respect to the MnSi crystal lattice. The reorientation is non-monotonic with increasing current, with the SkL rotating first in one direction and then the other. The SkL reorientation was studied at different sample locations and found to depend on the local current density  as inferred from a finite element analysis. The non-monotonic response indicates the presence of two competing effects on the SkL, most likely due to the presence of both radial electric and thermal currents. Such a scenario is supported by micromagnetic simulations, which show how these effects can act constructively or destructively to drive the SkL rotation, depending on the direction of the electric current. In addition, the simulations also suggest how the direction of the skyrmion flow may affect the SkL orientation.
\end{abstract}

\date{\today}

\maketitle

\section{Introduction}
\label{Introduction}
A magnetic skyrmion is a topologically protected spin texture, first discovered in MnSi in 2009~\cite{Muhlbauer:2009bc}.
In bulk materials, skyrmions arrange themselves into a periodic skyrmion lattice (SkL).
Skyrmions are typically found in magnetic materials with broken inversion symmetry where the Dzyaloshinskii-Moriya interaction (DMI) plays a critical role in stabilizing them~\cite{Dzyaloshinsky1958,Moriya1960,Nagaosa2013}.
In addition to MnSi, skyrmions have been observed in thin films of the helimagnet FeGe~\cite{Yu:2011hr}, the insulating multiferroic Cu$_2$OSeO$_3$~\cite{Seki:2012ie}, cubic $\beta$-Mn-type Co-Zn-Mn alloys~\cite{Tokunaga.2015}, and the polar magnetic semiconductor GaV$_4$S$_8$~\cite{Kezsmarki:2015bw} to name a few.
Skyrmions have recently been observed in centrosymmetric, but magnetically frustrated, materials where no DMI is present~\cite{Kurumaji2019, Khanh2020,Hirschberger2020, Casas2023}.

Skyrmions show promise for future data processing and storage devices due to their topological protection and the low current densities needed to manipulate them~\cite{Jonietz2010,Nagaosa2013,Okuyama2019,Back2020}.
While devices are likely to use individual skyrmions in thin film materials~\cite{Dupe2016,Wiesendanger2016,Fert2017,Gibertini2019}, developing a deeper understanding of the dynamics of the skyrmion lattice in bulk skyrmion materials not only provides valuable insights for optimizing these devices but also paves the way for exploring new possibilities for applications~\cite{Fert2013,Back2020}.
This is particularly true as excitations of individual skyrmions exhibit topologically non-trivial properties themselves~\cite{Schwarze2015}.

In bulk materials, small-angle neutron scattering (SANS) is a leading technique for probing the SkL~\cite{Muhlbauer2019}.
The SANS technique was also used for the first demonstration of skyrmion manipulation, where an electric current was found to cause an angular reorientation of the SkL in MnSi~\cite{Jonietz2010}.
Subsequently, similar effects were observed in insulating Cu$_2$OSeO$_3$ in response to an electric field~\cite{White2014}, and circularly polarized femtosecond laser pulses~\cite{Tengdin2022}.

Theory suggests that an electric current exerts two forces on the skyrmions: a drag force parallel to the current and a Magnus force perpendicular to it~\cite{Jonietz2010,Everschor2011,Everschor2012}.
Similarly, magnon currents generated by thermal gradients can drive skyrmion motion~\cite{Mochizuki:2014fja,Pollath2017, Yu2021}.
In a standard Hall bar geometry, this will lead to a translation of the SkL which is not observable by SANS.
However, a temperature gradient will cause a spatial variation in the spin-transfer torque, resulting in a torque on the SkL which lead to a rotation~\cite{Jonietz2010}.
Alternatively, edge friction was also reported to cause a rotation of the SkL~\cite{Okuyama2019}.
In both cases, the angular reorientation of the SkL required the presence of a secondary mechanism, be it a thermal gradient or an interaction with the sample edge.

\begin{figure}
    \includegraphics[width = 0.75\columnwidth]{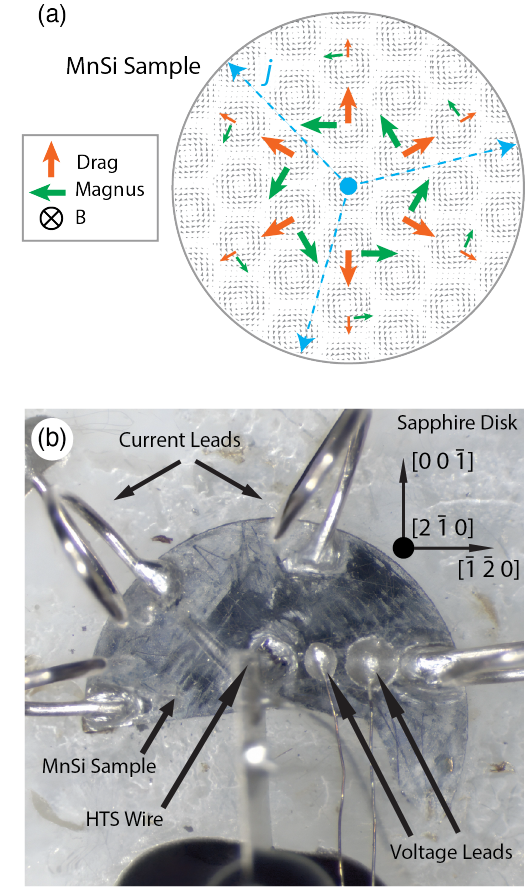}
    \caption{\label{ForceDiagramSample}
        (a) Schematic showing the drag (orange) and Magnus (green) forces acting on the skyrmions due to the electric current (blue) in an ideal Corbino geometry.
        A radially decreasing current density ($j \propto 1/r$) is formed in this geometry.
        A thermal gradient due to Joule heating will be induced and also decreases with increasing distance from the center.
        (b) (Half) Corbino sample used for the SANS experiment.
        }
\end{figure}

In this work, we study the SkL in MnSi under the influence of an electric current using a Corbino geometry.
Here, one current contact is placed at the center of a disc-shaped sample and another around the perimeter.
This generates a radial electric current through the sample, which decreases as $1/r$ with increasing distance ($r$) from the sample center.
As a result, the Magnus force will also vary as $1/r$, resulting in a torque on the SkL as illustrated in Fig.~\ref{ForceDiagramSample}(a).
Importantly, while a radial thermal gradient may arise due to Joule heating, it is not required for the manipulation of the SkL in this geometry.
Furthermore, the effects of edge friction are minimized as the skyrmions move primarily perpendicular to the circular sample edge.

We find that the SkL undergoes a not previously observed non-monotonic angular reorientation with respect to the MnSi crystal lattice in response to an increasing electric current.
As expected, the reorientation depends on the local radial electric current density, as confirmed by finite element analysis.
The non-monotonic response indicates the presence of multiple competing effects on the SkL, most likely due to the presence of a radial thermal gradient due to Joule heating.
This scenario is supported by micromagnetic simulations.

The paper is organized as follows. Sec. II provides the details of the samples, experimental setup, and measurements made. The results of the SANS measurements are then presented in Sec. III. Subsequently, Sec. IV shows the micromagnetic simulations of the system, identifying the key ingredients for the interpretation of the experimental data and corroborating the results of Sec. III. The mechanisms behind the observed results are then discussed on equal footing in Sec. V, before the concluding remarks are given in Sec. VI.

\section{Experimental Details}
\label{ExpDetails}
In this work, SANS was used to image the SkL.
This technique is particularly well suited to study the collective behavior of skyrmion lattices at mesoscopic length scales~\cite{Muhlbauer2019}, compared to direct space imaging techniques such as magnetic force microscopy (MFM) and Lorentz transmission electron microscopy (TEM)~\cite{Yu2010,Milde2013}.

The samples were prepared from a MnSi single crystal grown using the Bridgman method.
The growth was done by first melting stoichiometric amounts of Mn and Si in an alumina crucible within a Ta tube under UHP Ar at approximately 1550$^\circ$C. 
The resulting sample was then refined in a vertical Bridgman furnace with a 45$^\circ$C/cm gradient and a slow pass-through rate of 1.5$^\circ$C/hr to promote single-crystal growth. 
Once grown, the crystal was oriented using a Laue diffractometer.

The MnSi single crystal was cut into discs, using a diamond wire saw, with a surface normal along the $[2 \, \overline{1} \, 0]$ crystalline direction, a diameter of $6-8$~mm and a thickness of approximately $1.5$~ mm. 
After cutting, the discs were further thinned to a final thickness of $100 - 250$~$\mu$m by polishing and lapping.

The MnSi crystal was placed on a sapphire disc and current wires were attached by soldering.
To minimize Joule heating, a high-temperature superconducting (HTS) wire was used for the central current lead.
Silver wires were used for the current leads along the outer perimeter.
To reduce the likelihood of the sample cracking due to thermal cycling, the MnSi is not attached to the sapphire disc but is held in place only by the current leads.
Furthermore, the outer leads are wound into loops to provide strain relief.

Radial outward (positive) and inward (negative) electric currents with magnitudes up to $I = \pm 6$~A were applied during the SANS measurements.
To account for Joule heating and remain within the SkL hosting A-phase, radially placed voltage leads were used to measure the temperature-dependent resistance across the sample and maintain a constant average sample temperature.
A low current of 100 mA was used as a reference, and the set point of the cryomagnet temperature regulation was adjusted to keep the sample resistance constant and equal to $R_{\text{ref}} = 0.4186$~m$\Omega$ as the electric current was varied.
This corresponds to an average sample temperature of 29~K.

Preliminary SANS measurements were carried out at the D33 instrument at the Institut Laue-Langevin~\cite{5-42-540,5-42-568}.
All results presented here were obtained on the SANS-I instrument at the Paul Scherrer Institute, using the sample shown in Fig.~\ref{ForceDiagramSample}(b) with a thickness of ($130 \pm 4$)~$\mu$m and a radius of 3~mm.
This sample broke during the polishing and lapping but remains in a Corbino geometry with a $1/r$ current density in the region illuminated by the neutron beam as verified by a finite element analysis (FEA) of the current flow (see Appendix~\ref{FEA}).

\begin{figure*}
    \includegraphics{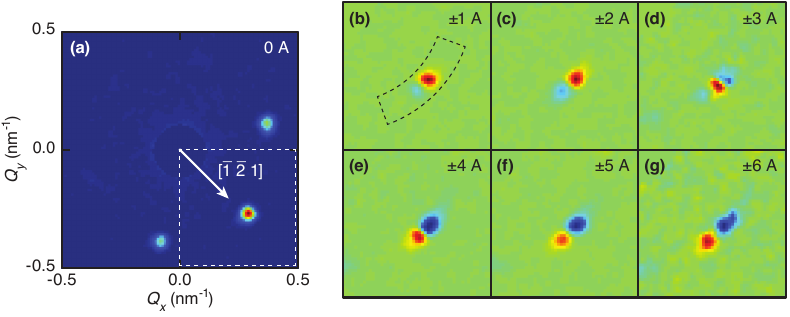}
    \caption{\label{DifPat}
        SANS SkL diffraction patterns averaged over the four sample locations indicated in the Fig.~\ref{IntDist}(b) inset.
        (a) Diffraction pattern obtained with no applied current.
        This is a sum of measurements as the bottom right peak is rocked through the Bragg condition.
        Background scattering near $Q = 0$ has been masked off.
        (b)--(g) Subtraction of diffraction patterns for positive and negative currents of the same magnitude and within the region of reciprocal space indicated by the dashed rectangle in (a).
        Here, red is positive, green is zero, and blue is negative intensity.
        The sector in (b) shows the region of reciprocal space included in Figs.~\ref{IntDist} and \ref{IntSub}.}
\end{figure*}

For the SANS experiment, the sample was placed in a horizontal field cryomagnet.
Horizontal and vertical translations of the magnet were used to control the region of the sample illuminated by the neutron beam, with the latter defined by a circular aperture with a diameter of 1~mm placed at the end of the beamline collimation section.
Restricting the beam aperture reduces both sample edge effects and the illumination of the current contacts and leads.
All reported measurements were obtained within a $2 \times 2$ ``pixel'' array located between the center and upper right current contacts as indicated in the Fig.~\ref{IntDist}(b) inset.

The SANS measurements were carried out with a neutron wavelength $\lambda_{\text{n}} = 0.6$~nm and bandwidth $\Delta \lambda_{\text{n}}/\lambda_{\text{n}} = 10\%$, a magnetic field of $0.22$~T applied normal to the sample surface and parallel to the incident neutron beam, and at a nominal temperature of 29~K. 
The magnetic field and temperature were chosen to maximize the SkL scattering signal, and are consistent with previous SANS studies of our MnSi samples~\cite{Leishman:2020tg}.
Before the application of a current, the sample was field-cooled from the paramagnetic phase.

\section{SANS Results}
\label{Results}
\begin{figure}
    \includegraphics{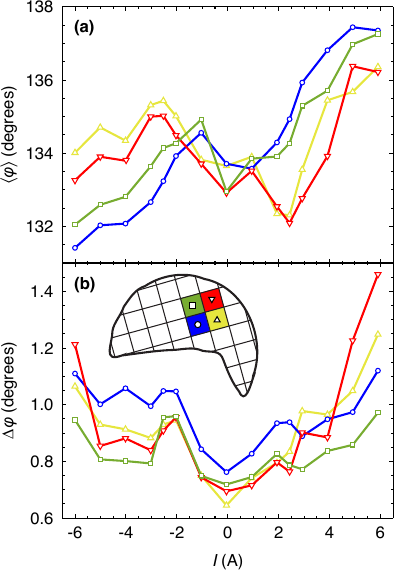}
    \caption{\label{IntDist}
        Spatially resolved azimuthal distribution of the SkL scattered intensity versus the current through the sample.
        (a) Center of mass-angle.
        (b) Standard deviation.
        Colors correspond to the four measurement locations within the Corbino sample, indicated in the inset.}
\end{figure}

The main experimental observations are illustrated in Fig.~\ref{DifPat}.
Without an applied current, the propagation vector of one of the hexagonal SkL Bragg peaks is seen to be aligned with the
$[\overline{1} \, \overline{2} \, 1]$ crystalline direction, as seen in Fig.~\ref{DifPat}(a).
To make efficient use of the neutron beam time, only one of the six equivalent SkL peaks is brought into the Bragg condition.
The remaining peaks therefore have a much reduced intensity or are not visible at all.
Once a current is applied, the SkL Bragg peaks rotate in opposite directions for positive and negative currents.
Figs.~\ref{DifPat}(b) -- \ref{DifPat}(g) show the subtraction of SkL diffraction patterns for currents of the same magnitude but opposite polarities.
For the two lowest currents, in Figs.~\ref{DifPat}(b) and \ref{DifPat}(c), an intensity ``dipole'' is observed with the intensity excess (red) rotated counterclockwise relative to the deficit (blue).
In contrast, the dipole is inverted at the three highest currents shown in Figs.~\ref{DifPat}(e) -- \ref{DifPat}(g).
At the intermediate current of $\pm 3$~A, there is an intensity ``quadrupole'' with a maximum in the middle and minima on both sides, as seen in Fig.~\ref{DifPat}(d).
The data shown in Fig.~\ref{DifPat} is an average of the four $\sim 1$~mm$^2$ locations of the sample indicated in the inset to Fig.~\ref{IntDist}(b).
Although there are differences from one location to another, which will be discussed below, they all exhibit qualitatively similar behavior.
The non-monotonic response to an increasing current suggests the presence of at least two competing effects on the SkL.

A quantitative, location-dependent measure of the SkL response is shown in Fig.~\ref{IntDist} for all the measured currents.
As the current is applied, the azimuthal intensity distribution broadens and is not well fitted by an analytical function such as a Gaussian or Lorentzian.
Therefore, the average angle
$\langle \varphi \rangle = \tfrac{1}{N} \sum_{i=1}^N \varphi_i$
and standard deviation
$\Delta \varphi = \tfrac{1}{N} \left( \sum_{i=1}^N ( \varphi_i - \langle \varphi \rangle )^2 \right)^{1/2}$
are used to characterize the SkL, shown in Figs.~\ref{IntDist}(a) and \ref{IntDist}(b) respectively.
The sums are over the $N$ neutrons detected within the angular section of reciprocal space indicated in Fig.~\ref{DifPat}(b), and $\varphi_i$ is the azimuthal angle on the $i$th neutron measured clockwise relative to the detector's positive $y$-axis.
The non-monotonic SkL rotation, which gives rise to the ``dipole'' switching in Figs.~\ref{DifPat}(b) -- \ref{DifPat}(g), is directly evident in the current dependence of $\langle \varphi \rangle$ in Fig.~\ref{IntDist}(a) for all four measurement locations. 
Here, a negative slope is seen for current magnitudes
$\left| I \right| \lesssim 2$~A and a positive slope is seen for
$\left| I \right| > 3$~A.
The azimuthal broadening of the SkL signal, shown in Fig.~\ref{IntDist}(b), is most pronounced for the higher currents.
The most likely origin of this broadening is a fracturing of the SkL into domains that rotate by varying amounts.

\begin{figure}
    \includegraphics{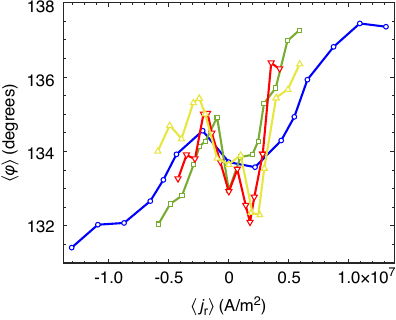}
    \caption{\label{IntDist2}
        Spatially resolved azimuthal distribution of the SkL scattered intensity versus the local radial current density.
        Colors correspond to the four measurement locations within the Corbino sample, indicated in the Fig.~\ref{IntDist}(b) inset.}
\end{figure}

If the SkL angular reorientation is current-driven, the radially decreasing current density provides a natural explanation for the azimuthal broadening seen in Fig.~\ref{IntDist}(b).
To directly compare the rotation observed at the different sample locations, Fig.~\ref{IntDist2} shows the SkL center-of-mass versus the local radial current density $\langle j_r \rangle$ obtained from the FEA (see Appendix~\ref{FEA}).
This causes most of the data from Fig.~\ref{IntDist}(a) to collapse onto a single curve, and notably the reversal in the rotation direction occurring at the same $\langle j_r \rangle \approx \pm 2 \times 10^6$~A/m$^2$.
This clearly demonstrates that the SkL rotation is directly related to the local current density in the sample.
The magnitude of $\langle \varphi \rangle$ is somewhat smaller within the reversal region for the innermost sample location.
This is likely due to the larger range of current densities, which will tend to average out the observed SkL rotation.
In addition, the SkL rotation may vary with temperature, with the innermost region of the sample being the warmest.

\begin{figure}
    \includegraphics{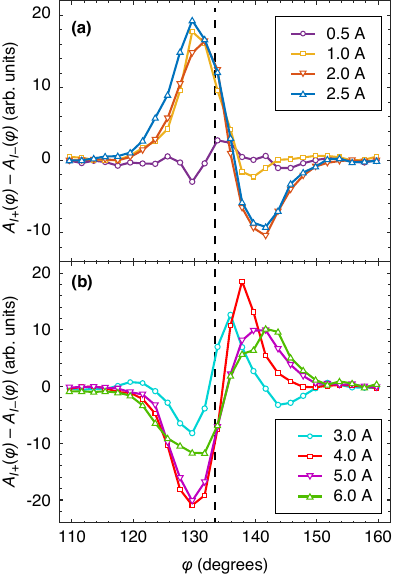}
    \caption{\label{IntSub}
        Azimuthal subtractions of the SkL scattered intensity for opposite currents within the region of reciprocal space indicated in Fig.~\ref{DifPat}(b).
        Data is the average over the yellow (up triangle) and green (square) sample locations indicated in the Fig.~\ref{IntDist}(b) inset.
        (a) Low currents: $0.5$~A~$-$~$2.5$~A. 
        (b) High currents: 3~A~$-~$6~A.
        The vertical dashed line indicates the average SkL peak position for 0~A.
        }
\end{figure}

Figure~\ref{IntSub} shows the azimuthal intensity distribution for opposite current subtractions, $A_{I+}(\varphi) - A_{I-}(\varphi)$.
Only data at the two pixels located at the same radial distance from the center contact are included (green/square and yellow/up triangle pixels in Fig.~\ref{IntDist}(b)), as they experience the same current density.
For all currents, the midpoint of the intensity dipole is rotated roughly $4^{\circ}$ degrees clockwise (higher $\varphi$) relative to the location of the SkL Bragg peak without an applied current which is indicated by the vertical dashed line.
Below the cross-over, shown in Fig~\ref{IntSub}(a), the locations of the maxima and minima are independent of the current magnitude.
However, the positive amplitude at $\varphi \sim 132^{\circ}$ is roughly two times greater than the negative amplitude at $\varphi \sim 140^{\circ}$.
Above the cross-over, in Fig.~\ref{IntSub}(b), positive and negative peak amplitudes are comparable, where the location of the maximum moves to higher angles with increasing current.

The asymmetry, both with respect to the rotation direction and the intensity maxima and minima, underscores the presence of competing effects that can only partially offset each other.
We speculate that the most likely origin is the competition between electric and thermal currents, with the latter arising from Joule heating in the sample.
Although the direction of the electric current is reversible, the thermal gradient will always be directed from the edge of the sample towards the center, resulting in a non-monotonic SkL rotation behavior.
These effects likely have a different dependence on the electric current magnitude.

\section{Micromagnetic Simulations}
\label{Simulations}
Micromagnetic simulations were performed to complement the SANS data and test the hypothesis of competing electric and thermal effects.
These used the Mumax-3.10 package~\cite{Vansteenkiste2014}, for a ferromagnetic film of the same shape as that used for the SANS measurements although with a much smaller 8~$\mu$m diameter, discretized in micromagnetic cells of size $4 \times 4 \times4$~nm$^3$. 
Due to computational limitations, it is not possible to perform micromagnetic simulations of the system at its actual physical size.
However, the simulations are expected to be valid as the radial current density and thermal gradient profiles, which dictate bulk behavior, are preserved across length scales.
The following parameters were used: saturation magnetization $M_{\text{S}} = 1.52 \times 10^5$~A/m, exchange stiffness $A_{\text{ex}} = 3.4 \times 10^{-3}$~J/m, and bulk Dzialoshinkii-Morya strength $D = 2.1 \times 10^{-3}$~J/m$^2$~\cite{Karhu2012}.

The system is subjected to an applied magnetic field of $0.22$~T, and periodic boundary conditions are assumed along the field axis.
A radial spin-polarized current is applied from the center to the perimeter of the sample, with a distribution obtained by solving Poisson’s equation in the micromagnetic framework for an applied voltage difference $\Delta U$ (see Appendix~\ref{PoissonSolver}).
The spin current polarization is set to $P = 0.1$~\cite{Neubauer2009,Jonietz2010}, and a material electric conductivity of $\sigma = 3.9 \times 10^6$~S/m~\cite{Stishov2008} and a Gilbert damping parameter of $\alpha = 0.01$ were used. 

\begin{figure}
    \includegraphics{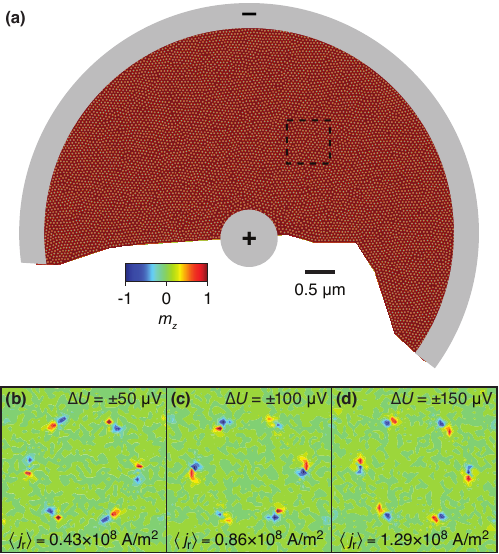}
    \caption{\label{SimSkL}
        SkL angular reorientation obtained from micromagnetic simulations.
        (a) Sample geometry used in the simulations.
        Colors indicate the sample magnetization with $m_z = -1$ corresponding to the skyrmion centers.
        Current contacts, shown in grey, are placed at the center and perimeter of the sample.
        (b)-(c) SkL structure factor subtractions, $S_{I+}({\bf Q}) - S_{I-}({\bf Q})$, for a low, intermediate and high current respectively.
        The average current density $\langle j_r \rangle$ is indicated for each case and $S({\bf Q})$ is calculated for the region of the sample indicated by the dashed box in (a).
        In all cases, the structure factor was computed after the system had evolved for 4~$\mu$s to achieve a steady state configuration.
        }
\end{figure}

The temperature gradient is assumed to follow a $1/r$ dependence, where $r$ is the distance from the central current contact, with $\Delta T=5$~K across the whole sample and $\Delta T\approx 0.2$~K across the investigated region indicated in Fig.~\ref{SimSkL}(a). 
The temperature profile is kept constant for all magnitudes of applied current.
Given the relatively narrow temperature range and focus on a small region of the sample, the changes in $P$ due to the thermal gradient are negligible compared to the $1/r$ dependence of the current density and thus ignored in the simulations.

\begin{figure}
    \includegraphics{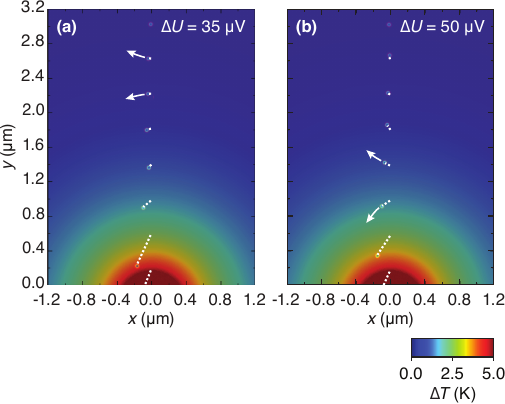}
    \caption{\label{SimSingle}
        Competition between single-skyrmion forces due to the electric current and the thermal gradient obtained from micromagnetic simulations.
        Colors indicate the local temperature with the hot region near the central contact located at the bottom, and the electric current is along the vertical direction towards the top.
        Dotted lines indicate skyrmion trajectories for different starting positions along the radial direction ($x = 0$) for (a) a lower current and (b) a higher current. In both cases, simulations were performed for 2~$\mu$s.
        }
\end{figure}

Before investigating the collective dynamics of the SkL, the behavior of a single skyrmion was studied.
Both an electric current and a thermal gradient will independently give rise to both a radial force and a perpendicular Magnus force~\cite{Pollath2017, Yu2021}.
Here, the thermal gradient results in a radial force experienced by the skyrmions towards the high-temperature region at the sample center~\cite{raimondo2022temperature}.
When both an electric current and a thermal gradient are present, the resultant force experienced by the skyrmion reflects the competition between the two effects.
The net force on the skyrmion depends on the local temperature and current density and can lead to an inward or outward motion, as shown in Fig.~\ref{SimSingle}.

For the lower current in Fig.~\ref{SimSingle}(a), the radial components of the forces balance each other at a vertical distance of $y \approx 2.4$~$\mu$m.
For the higher current in Fig.~\ref{SimSingle}(b) this moves closer to the center at $y \approx 1.2$~$\mu$m.

Simulations of the collective skyrmion system under the influence of an electric current and a thermal gradient allow for a direct comparison to the SANS results. 
Figure~\ref{SimSkL}~(b)-(d) shows subtractions of the SkL structure factor,
$S({\bf Q}) \propto \left| \sum_{i=1}^{N} \exp (-i {\bf Q} \cdot {\bf r}_i) \right|^2$,
obtained from simulations with a positive and negative applied voltage difference (current) of the same magnitude.
Here, ${\bf r}_i$ is the position of the $i$th skyrmion, and the summation runs over the $N$ skyrmions within the region of the sample indicated in Fig.~\ref{SimSkL}~(a) which roughly corresponds to the region illuminated in the SANS experiment.
The resemblance to the diffraction patterns in Fig.~\ref{DifPat} is striking, with a non-monotonic rotation between the two directions of the current and intensity dipoles that switch direction as the current magnitude is increased.

\begin{figure}
    \includegraphics{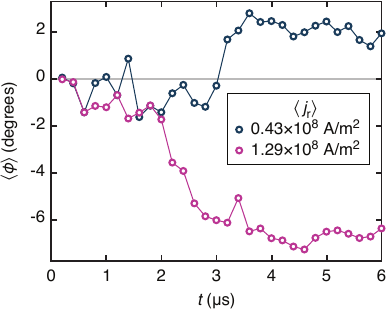}
    \caption{\label{TimeDep}
        Time dependence of the average SkL orientation angle $\langle \phi \rangle$, obtained from the bond-orientational order parameter $\psi_6$, and two different values of $\langle j_r \rangle$.
        }
\end{figure}

The micromagnetic simulations provide both local and time-resolved information about the SkL which cannot be readily obtained from the SANS measurements. 
An example is the bond-orientational order parameter,
$\psi_6 = \frac{1}{N_b}\sum_{i=1}^{N_b} \exp \left( 6i \theta_{ij} \right)$,
where $N_b$ is the number of nearest neighbors (NN) of the $i$th skyrmion and $\theta_{ij}$ the bond angle between this skyrmion and its $j$th NN~\cite{Menezes2017}.
This yields the mean orientation at the local lattice site centered at the $i^{th}$ skyrmion, which, when averaged over all the skyrmions, corresponds to the mean SkL orientation:
$\langle \phi \rangle = \left\langle \frac{1}{6} \arctan \left( \frac{\operatorname{Im}(\psi_6)}{\operatorname{Re}(\psi_6)} \right) \right\rangle$.
Figure~\ref{TimeDep} shows $\langle \phi \rangle$ obtained from the micromagnetic simulations as a function of time, and currents corresponding to Figs.~\ref{SimSkL}(b) and \ref{SimSkL}(d) and the same region of the sample.
This shows that when the currents are applied, the SkL goes through a transient period of lattice rotation before reaching a stable angular orientation after $\sim 4~\mu$s.
One should note that even in this steady-state configuration for the SkL orientation, the skyrmions are constantly moving due to the electric current and thermal gradient.

In addition to the effects already discussed, the skyrmion motion itself may also influence the SkL orientation in analogy with the superconducting vortex lattice~\cite{Braun:1996tn}.
Focusing on the region of the sample indicated in Fig.~\ref{SimSkL}~(a), one can resolve how the skyrmion flow is related to the SkL rotation.
\begin{figure*}
    \includegraphics{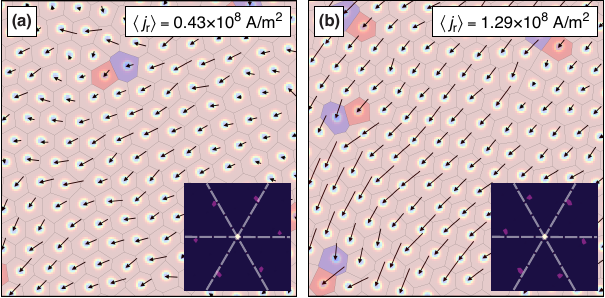}
    \caption{\label{SimFlow}
        Skyrmion flow in a low (a) and high (b) positive current regime, for the region of the sample indicated in Fig.~\ref{SimSkL}(a).
        The arrows indicate the skyrmion velocity after the system has evolved for 5~$\mu$s (i.e., 1~$\mu$s after the SkL orientation has stabilized).
        Lattice dislocations, consisting of bound five- and sevenfold coordinated skyrmions, are indicated by red and blue unit cells respectively.
        Insets show the corresponding SkL structure factor, $S({\bf Q})$.
        }
\end{figure*}
In the regime of low current, the SkL rotation originates from the temperature and current gradients, where the forces experienced by the skyrmions vary in strength across a SkL domain. This can be described
by a continuous model and considering a rigid SkL~\cite{Everschor2011,Everschor2012}. 
In this case, skyrmion flow is strongly affected by the lattice orientation, due to skyrmion-skyrmion repulsion, as shown in Fig.~\ref{SimFlow}(a). 
In contrast, in the regime of high currents, where the inversion of the SkL rotations is observed, a realignment of the lattice planes is found to facilitate the flow of skyrmions along the current direction, as shown in Fig.~\ref{SimFlow}~(b).
Such realignment of the SkL is possible due to the formation of lattice dislocations, which release the shear tension of the SkL. 
The insets in Fig.~\ref{SimFlow} show the corresponding SkL structure factors, where the inversion of the SkL rotation is observed when the current is increased.

\section{Discussion}
\label{Discussion}
As demonstrated by the SANS measurements and micromagnetic simulations, the response of the SkL subjected to an external drive is affected by several mechanisms.
These include effects of both the electric current and the resulting thermal gradient due to Joule heating, as well as a preferred SkL orientation relative to the direction of skyrmion motion.
Together, these give rise to the non-monotonic angular reorientation of the SkL.

The choice of the Corbino geometry allowed both the radially decaying electric current and thermal gradient to independently induce rotations of the SkL.
In contrast, the Hall bar geometry relies on the thermal gradient to provide the necessary spin transfer torque gradient. 
While the Corbino geometry introduces a more complex interplay of effects, it gives rise to the observed non-monotonic behavior, offering new insights into competing mechanisms within the system.

The combination of multiple effects makes it difficult to experimentally separate the influence of each independently.
However, it is possible to study the effects of an isolated electric current or thermal gradient in micromagnetic simulations, as shown in the Appendix~\ref{SeparateElThCurrent}.
This reinforces our conclusion that the dipole switching, observed when both effects are present, likely results from the competition between the two effects.

We note that while the simulations, for the sake of simplicity, assumed a thermal gradient that followed the same $1/r$ dependence as the electric current, this is most likely not accurate.
Although still radially decaying, the thermal profile will also depend on the coupling between the sample and the environment, including the surrounding exchange gas.
In contrast to the electric current density, the exact thermal profile is therefore not easily obtained from an FEA.
Despite this limitation, the ability of the simulations to reproduce the SANS results is remarkable.

In addition to the effects discussed above, two additional conclusions may be drawn from our studies.
First, only an angular reorientation is observed, rather than a continuously rotating SkL.
In the absence of a drive, the SkL is aligned along the crystalline directions corresponding to the minimum in the angular dependence of the free energy.
The applied electric current and the associated thermal gradient produce a net torque on the SkL and cause an angular reorientation.
The magnitude of the reorientation corresponds to the angle at which the torque is balanced by the gradient in the angular free energy.
Clearly, the torque is not sufficient to overcome the free energy barrier that separates the minima, each $60^{\circ}$, and produce a freely rotating SkL.
This may be due to the electric and thermal effects offsetting each other.
Second, no change to the magnitude of the SkL scattering vector $\left| \bf{Q} \right|$ is observed as a function of the applied current or sample location.
Hence, the SkL periodicity, and thus the skyrmion density, remains constant within the precision of our SANS measurements.
Therefore, skyrmions must be continuously nucleated and annihilated at the center and perimeter of the sample.

Finally, our results indicate that it should be possible to manipulate the SkL with a pure thermal gradient.
This has indeed been observed for Cu$_2$OSeO$_3$ using x-ray diffraction~\cite{Zhang:2018bg,Jin.2024ajk}, although the interpretation of these measurements is complicated as the x-rays themselves will cause a local heating of the sample.
Skyrmion motion induced by heat flow was also observed in real space by Lorentz Transmission Electron Microscopy~\cite{Mochizuki:2014fja, Yu2021}.

\section{Conclusion}
\label{Conclusion}
In this work, we successfully demonstrated how a radial electric current and thermal gradient lead to a nontrivial angular reorientation of the SkL in MnSi using SANS.
This is, to our knowledge, the first observation of such a non-monotonic SkL response to an external drive.
Measurements at different sample locations showed conclusively that the amplitude of the SkL rotation depends on the local current density.
The measurements were complemented by micromagnetic simulations which show how the electric current and thermal gradient may act constructively or destructively to drive the SkL rotation, depending on the direction of the electric current.
Finally, the direction of the skyrmion flow may also affect the SkL orientation.
These effects will have to be managed carefully in any skyrmionic device and will contribute favorably to the nonlinear response needed for the further use of skyrmion lattices in emergent unconventional computing \cite{yokouchi2022pattern,li2021magnetic}.

\appendix

\section{Finite Element Analysis}
\label{FEA}
\begin{figure}
    \includegraphics{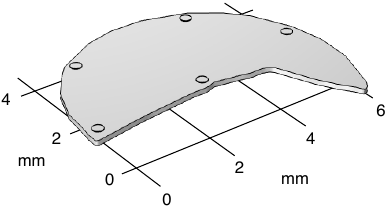}
    \caption{\label{FEAsample}
        Finite element analysis model of the sample used for the SANS measurements.}
\end{figure}

To determine the degree to which the current density follows the $1/r$-dependence expected for an ideal Corbino geometry, a finite element analysis (FEA) of the SANS sample was performed using the COMSOL Multiphysics\textsuperscript{\textregistered} software package \cite{comsol}.
The sample geometry used for the FEA is shown in Fig.~\ref{FEAsample}.
The following MnSi parameters extracted at 29~K were used:
specific heat at constant pressure $C_{\text{P}} = 2.86$~J/kg\,K~\cite{Cheng2010},
thermal conductivity $\kappa = 5.75$~W/m\,K~\cite{Hirokane2016},
and electric conductivity $\sigma = 3.9 \times 10^6$~S/m~\cite{Stishov2008}.

\begin{figure}
    \includegraphics{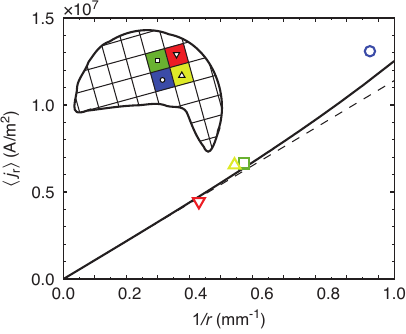}
    \caption{\label{FEAjvsr}
        Radial current versus radial distance was obtained from a finite element analysis and averaged over the four MnSi sample locations illuminated during the SANS measurement (open symbols).
        The dashed line shows the $1/r$-behavior expected for an ideal Corbino geometry and the solid line is averaged over a 1~mm radial width (see text for details).}
\end{figure}

Figure~\ref{FEAjvsr} shows the average radial current density $\langle j_{\text{r}}(r) \rangle$ obtained from the FEA versus inverse distance from the center contact for each of the four SANS measurement locations.
Due to the rapidly changing current density which skews the results, this deviates from the purely linear behavior expected for an ideal Corbino geometry indicated by the dashed line.
The deviation is most prominent close to the central contact (largest values of $1/r$).
Averaging the ideal $1/r$-dependence over a radial width $a$ yields $j_0 \int_{r-a/2}^{r+a/2} \, (1/r) \, dr = j_0 \ln (\tfrac{r+a/2}{r-a/2})$, indicated by the solid line in Fig.~\ref{FEAjvsr}.
Here, $j_0$ is an adjustable scaling parameter.
The FEA results are in good agreement with the expected behavior, in particular for the three measurement locations furthest from the sample center.
From this, we conclude that the current distribution in the samples used for the SANS experiment corresponds to a Corbino geometry, even though it was made using an incomplete disc and with discrete contacts at the perimeter rather than a continuous one. 
Furthermore, it is possible to relate the SkL angular reorientation to the local radial current density, as shown in Fig.~\ref{IntDist2}.

\section{Micromagnetic Framework}
\label{PoissonSolver}
Within the micromagnetic framework, we describe the magnetization of the sample by considering the vector field
${\bf M}({\bf r}) = M_{\text{s}} \, {\bf m}({\bf r})$
with constant magnetization modulus
$|{\bf M}|= M_{\text{s}}$
and the normalized magnetization direction
${\bf m}({\bf r})$
at each point ${\bf r} \in \mathbb{R}^2$ of the film.
Periodic boundary conditions are assumed along the magnetic field axis, perpendicular to the film plane.
The dynamics of the magnetization are governed by the Landau-Lifshitz-Gilbert (LLG) equation
$\dot{{\bf m}} = - \gamma {\bf m} \times {\bf H}_{\text{eff}} + \alpha {\bf m}\times\dot{{\bf m}} + {\bm \tau}_{\text{STT}}$,
where $\gamma$ is the gyromagnetic ratio, $\alpha$ the dimensionless damping factor, and
${\bf H}_{\text{eff}}$
the effective field, which can be derived from the magnetic free energy~$E({\bf m})$ by taking the functional derivative with respect to the magnetization:
${\bf H}_{\text{eff}} = - \delta E/\delta {\bf M}$.
Note that, in the presence of an applied current density ${\bf j}$, we extend the LLG equation by adding the torque~${\bm \tau}_{\text{STT}}$ which includes the adiabatic and non-adiabatic spin-transfer-torque (STT) terms derived by Zhang and Li~\cite{Zhang2004}:
${\bm \tau}_{\text{STT}} = - {\bf m} \times({\bf m} \times ({\bf v} \cdot \nabla) {\bf m}) + \beta {\bf m} \times ({\bf v} \cdot \nabla) {\bf m}$,
where ${\bf v} = -\frac{\mu_{\text{B}} P}{e M_{\text{s}} (1+\beta^2)} {\bf j}$.
Here, $\nabla$ is the two-dimensional differential operator, $\beta$ is a dimensionless constant that represents the degree of non-adiabaticity, $e$ the elementary charge, $\mu_{\text{B}}$ the Bohr magneton, and $P$ is the spin-current polarization.

\begin{figure}
    \includegraphics{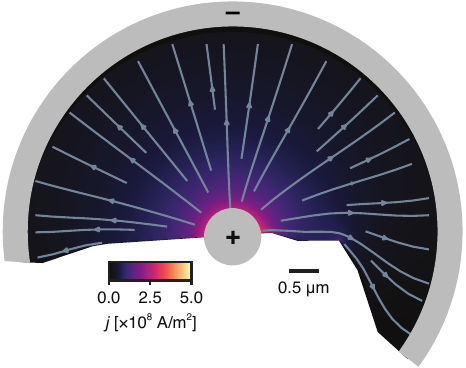}
    \caption{\label{Poison}
        Current distribution obtained by solving the Poisson equation for $\Delta U = 50$~$\mu$V and a uniform out-of-plane ferromagnetic state.
        Voltage contacts, shown in grey, are placed at the center and perimeter of the sample.
        } 
\end{figure}

To simulate the SkL under nonuniform current distributions, a Poisson solver was implemented in the micromagnetic simulations, where a custom module has been added to the simulation package Mumax-$3.10$~\cite{Vansteenkiste2014}.
Considering a set of ``contact'' micromagnetic cells where an electric potential $u$ is fixed, the generalized Poisson equation $\nabla \cdot (\sigma\nabla u) = 0 $ is solved using a conjugate gradient method for the considered geometry, from where the current density ${\bf j}=-\sigma \nabla u$ is calculated.
Figure~\ref{Poison} shows an example of the calculated current distribution.
More details on the method are provided in Ref.~\onlinecite{Prychynenko2018}.

\section{Separate Effects of Electric Currents and Thermal Gradients}
\label{SeparateElThCurrent}
\begin{figure}
    \includegraphics{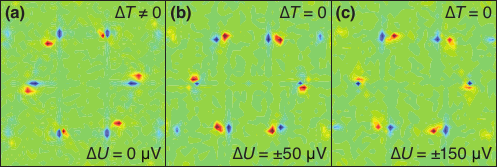}
    \caption{\label{ThElGradient}
        SkL structure factor subtractions obtained from micromagnetic simulations for separate temperature or current density gradients.
        (a) Only a thermal gradient with no electric current.
        Here, $\Delta T=5$~K across the whole sample and $\Delta T\approx 0.2$~K across the investigated region indicated in Fig.~\ref{SimSkL}~(a).
        The subtraction is done with respect to the SkL configuration with no temperature gradient,
        $S_{\Delta T \neq 0}({\bf Q}) - S_{\Delta T = 0}({\bf Q})$.
        (b,c) Low and high electric current regimes and no thermal gradient.
        Here, the subtraction is for opposite polarities of the electric current, $S_{I+}({\bf Q}) - S_{I-}({\bf Q})$.
        In all cases, the structure factor was computed after the system had evolved for 4~$\mu$s.}    
\end{figure}

Experimentally, it is not possible to study the SkL under the influence of only an electric current or a thermal gradient.
However, micromagnetic simulations of the SkL dynamics allow one to investigate the separate effects of the two driving mechanisms on the SkL reorientation,  as shown in Fig.~\ref{ThElGradient}.
This shows that both gradients in the temperature and current density will independently cause an angular reorientation of the SkL, which, depending on the direction of the electric current, can be in the same or opposite directions.
This reinforces the conclusion that the dipole switching, observed when both effects are present, results from the competition between the electric and thermal effects.

\begin{acknowledgments}
This work was supported by the U.S. Department of Energy, Office of Basic Energy Sciences, under Award No. DE-SC0005051 (N.C., A.W.D.L., G.L., M.R.E.) and the Research Foundation - Flanders (FWO-Vlaanderen) and Brazilian Agencies FACEPE, CAPES and CNPq (R.M.M., M.V.M.). 
Crystal growth at Los Alamos National Laboratory was performed under the auspices of the U.S. Department of Energy, Office of Basic Energy Sciences, Division of Materials Science and Engineering under the project “Quantum Fluctuations in Narrow-Band Systems (Y.L., E.D.B.).
Sample mounting and magneto-transport measurements at Argonne National Laboratory were supported by the U. S. Department of Energy, Office of Science, Basic Energy Sciences, Materials Sciences and Engineering Division (U.W, W.K.K).
This work is based on experiments performed at the Swiss Spallation Neutron Source SINQ, Paul Scherrer Institute, Villigen, Switzerland (J.S.W., M.B., M.J.).
Part of this work is based on experiments performed at the Institut Laue-Langevin, Grenoble, France (R.C.).
We are grateful to A.~Oliver for help in aligning the MnSi single crystal, and to T.~Wen for assistance with the finite element analysis.
\end{acknowledgments}

\bibliography{MnSiCorbino.bib}

\begin{thebibliography}{50}%
\makeatletter
\providecommand \@ifxundefined [1]{%
 \@ifx{#1\undefined}
}%
\providecommand \@ifnum [1]{%
 \ifnum #1\expandafter \@firstoftwo
 \else \expandafter \@secondoftwo
 \fi
}%
\providecommand \@ifx [1]{%
 \ifx #1\expandafter \@firstoftwo
 \else \expandafter \@secondoftwo
 \fi
}%
\providecommand \natexlab [1]{#1}%
\providecommand \enquote  [1]{``#1''}%
\providecommand \bibnamefont  [1]{#1}%
\providecommand \bibfnamefont [1]{#1}%
\providecommand \citenamefont [1]{#1}%
\providecommand \href@noop [0]{\@secondoftwo}%
\providecommand \href [0]{\begingroup \@sanitize@url \@href}%
\providecommand \@href[1]{\@@startlink{#1}\@@href}%
\providecommand \@@href[1]{\endgroup#1\@@endlink}%
\providecommand \@sanitize@url [0]{\catcode `\\12\catcode `\$12\catcode
  `\&12\catcode `\#12\catcode `\^12\catcode `\_12\catcode `\%12\relax}%
\providecommand \@@startlink[1]{}%
\providecommand \@@endlink[0]{}%
\providecommand \url  [0]{\begingroup\@sanitize@url \@url }%
\providecommand \@url [1]{\endgroup\@href {#1}{\urlprefix }}%
\providecommand \urlprefix  [0]{URL }%
\providecommand \Eprint [0]{\href }%
\providecommand \doibase [0]{https://doi.org/}%
\providecommand \selectlanguage [0]{\@gobble}%
\providecommand \bibinfo  [0]{\@secondoftwo}%
\providecommand \bibfield  [0]{\@secondoftwo}%
\providecommand \translation [1]{[#1]}%
\providecommand \BibitemOpen [0]{}%
\providecommand \bibitemStop [0]{}%
\providecommand \bibitemNoStop [0]{.\EOS\space}%
\providecommand \EOS [0]{\spacefactor3000\relax}%
\providecommand \BibitemShut  [1]{\csname bibitem#1\endcsname}%
\let\auto@bib@innerbib\@empty
\bibitem [{\citenamefont {Mühlbauer}\ \emph {et~al.}(2009)\citenamefont
  {Mühlbauer}, \citenamefont {Binz}, \citenamefont {Jonietz}, \citenamefont
  {Pfleiderer}, \citenamefont {Rosch}, \citenamefont {Neubauer}, \citenamefont
  {Georgii},\ and\ \citenamefont {Böni}}]{Muhlbauer:2009bc}%
  \BibitemOpen
  \bibfield  {author} {\bibinfo {author} {\bibfnamefont {S.}~\bibnamefont
  {Mühlbauer}}, \bibinfo {author} {\bibfnamefont {B.}~\bibnamefont {Binz}},
  \bibinfo {author} {\bibfnamefont {F.}~\bibnamefont {Jonietz}}, \bibinfo
  {author} {\bibfnamefont {C.}~\bibnamefont {Pfleiderer}}, \bibinfo {author}
  {\bibfnamefont {A.}~\bibnamefont {Rosch}}, \bibinfo {author} {\bibfnamefont
  {A.}~\bibnamefont {Neubauer}}, \bibinfo {author} {\bibfnamefont
  {R.}~\bibnamefont {Georgii}},\ and\ \bibinfo {author} {\bibfnamefont
  {P.}~\bibnamefont {Böni}},\ }\bibfield  {title} {\bibinfo {title} {{Skyrmion
  Lattice in a Chiral Magnet}},\ }\href
  {https://doi.org/10.1126/science.1166767} {\bibfield  {journal} {\bibinfo
  {journal} {Science}\ }\textbf {\bibinfo {volume} {323}},\ \bibinfo {pages}
  {915} (\bibinfo {year} {2009})}\BibitemShut {NoStop}%
\bibitem [{\citenamefont {Dzyaloshinsky}(1958)}]{Dzyaloshinsky1958}%
  \BibitemOpen
  \bibfield  {author} {\bibinfo {author} {\bibfnamefont {I.}~\bibnamefont
  {Dzyaloshinsky}},\ }\bibfield  {title} {\bibinfo {title} {{A thermodynamic
  theory of “weak” ferromagnetism of antiferromagnetics}},\ }\href
  {https://doi.org/10.1016/0022-3697(58)90076-3} {\bibfield  {journal}
  {\bibinfo  {journal} {J. Phys. Chem. Solids}\ }\textbf {\bibinfo {volume}
  {4}},\ \bibinfo {pages} {241} (\bibinfo {year} {1958})}\BibitemShut {NoStop}%
\bibitem [{\citenamefont {Moriya}(1960)}]{Moriya1960}%
  \BibitemOpen
  \bibfield  {author} {\bibinfo {author} {\bibfnamefont {T.}~\bibnamefont
  {Moriya}},\ }\bibfield  {title} {\bibinfo {title} {{Anisotropic Superexchange
  Interaction and Weak Ferromagnetism}},\ }\href
  {https://doi.org/10.1103/PhysRev.120.91} {\bibfield  {journal} {\bibinfo
  {journal} {Phys. Rev.}\ }\textbf {\bibinfo {volume} {120}},\ \bibinfo {pages}
  {91} (\bibinfo {year} {1960})}\BibitemShut {NoStop}%
\bibitem [{\citenamefont {Nagaosa}\ and\ \citenamefont
  {Tokura}(2013)}]{Nagaosa2013}%
  \BibitemOpen
  \bibfield  {author} {\bibinfo {author} {\bibfnamefont {N.}~\bibnamefont
  {Nagaosa}}\ and\ \bibinfo {author} {\bibfnamefont {Y.}~\bibnamefont
  {Tokura}},\ }\bibfield  {title} {\bibinfo {title} {{Topological properties
  and dynamics of magnetic skyrmions}},\ }\href
  {https://doi.org/10.1038/nnano.2013.243} {\bibfield  {journal} {\bibinfo
  {journal} {Nat. Nanotechnol.}\ }\textbf {\bibinfo {volume} {8}},\ \bibinfo
  {pages} {899–911} (\bibinfo {year} {2013})}\BibitemShut {NoStop}%
\bibitem [{\citenamefont {Yu}\ \emph {et~al.}(2010{\natexlab{a}})\citenamefont
  {Yu}, \citenamefont {Kanazawa}, \citenamefont {Onose}, \citenamefont
  {Kimoto}, \citenamefont {Zhang}, \citenamefont {Ishiwata}, \citenamefont
  {Matsui},\ and\ \citenamefont {Tokura}}]{Yu:2011hr}%
  \BibitemOpen
  \bibfield  {author} {\bibinfo {author} {\bibfnamefont {X.~Z.}\ \bibnamefont
  {Yu}}, \bibinfo {author} {\bibfnamefont {N.}~\bibnamefont {Kanazawa}},
  \bibinfo {author} {\bibfnamefont {Y.}~\bibnamefont {Onose}}, \bibinfo
  {author} {\bibfnamefont {K.}~\bibnamefont {Kimoto}}, \bibinfo {author}
  {\bibfnamefont {W.~Z.}\ \bibnamefont {Zhang}}, \bibinfo {author}
  {\bibfnamefont {S.}~\bibnamefont {Ishiwata}}, \bibinfo {author}
  {\bibfnamefont {Y.}~\bibnamefont {Matsui}},\ and\ \bibinfo {author}
  {\bibfnamefont {Y.}~\bibnamefont {Tokura}},\ }\bibfield  {title} {\bibinfo
  {title} {{Near room-temperature formation of a skyrmion crystal in thin-films
  of the helimagnet FeGe}},\ }\href {https://doi.org/10.1038/nmat2916}
  {\bibfield  {journal} {\bibinfo  {journal} {Nat. Mater.}\ }\textbf {\bibinfo
  {volume} {10}},\ \bibinfo {pages} {106} (\bibinfo {year}
  {2010}{\natexlab{a}})}\BibitemShut {NoStop}%
\bibitem [{\citenamefont {Seki}\ \emph {et~al.}(2012)\citenamefont {Seki},
  \citenamefont {Yu}, \citenamefont {Ishiwata},\ and\ \citenamefont
  {Tokura}}]{Seki:2012ie}%
  \BibitemOpen
  \bibfield  {author} {\bibinfo {author} {\bibfnamefont {S.}~\bibnamefont
  {Seki}}, \bibinfo {author} {\bibfnamefont {X.~Z.}\ \bibnamefont {Yu}},
  \bibinfo {author} {\bibfnamefont {S.}~\bibnamefont {Ishiwata}},\ and\
  \bibinfo {author} {\bibfnamefont {Y.}~\bibnamefont {Tokura}},\ }\bibfield
  {title} {\bibinfo {title} {{Observation of Skyrmions in a Multiferroic
  Material}},\ }\href {https://doi.org/10.1126/science.1214143} {\bibfield
  {journal} {\bibinfo  {journal} {Science}\ }\textbf {\bibinfo {volume}
  {336}},\ \bibinfo {pages} {198} (\bibinfo {year} {2012})}\BibitemShut
  {NoStop}%
\bibitem [{\citenamefont {Tokunaga}\ \emph {et~al.}(2015)\citenamefont
  {Tokunaga}, \citenamefont {Yu}, \citenamefont {White}, \citenamefont
  {R{\o}nnow}, \citenamefont {Morikawa}, \citenamefont {Taguchi},\ and\
  \citenamefont {Tokura}}]{Tokunaga.2015}%
  \BibitemOpen
  \bibfield  {author} {\bibinfo {author} {\bibfnamefont {Y.}~\bibnamefont
  {Tokunaga}}, \bibinfo {author} {\bibfnamefont {X.~Z.}\ \bibnamefont {Yu}},
  \bibinfo {author} {\bibfnamefont {J.~S.}\ \bibnamefont {White}}, \bibinfo
  {author} {\bibfnamefont {H.~M.}\ \bibnamefont {R{\o}nnow}}, \bibinfo {author}
  {\bibfnamefont {D.}~\bibnamefont {Morikawa}}, \bibinfo {author}
  {\bibfnamefont {Y.}~\bibnamefont {Taguchi}},\ and\ \bibinfo {author}
  {\bibfnamefont {Y.}~\bibnamefont {Tokura}},\ }\bibfield  {title} {\bibinfo
  {title} {{A new class of chiral materials hosting magnetic skyrmions beyond
  room temperature}},\ }\href {https://doi.org/10.1038/ncomms8638} {\bibfield
  {journal} {\bibinfo  {journal} {Nat. Commun.}\ }\textbf {\bibinfo {volume}
  {6}},\ \bibinfo {pages} {7638} (\bibinfo {year} {2015})}\BibitemShut
  {NoStop}%
\bibitem [{\citenamefont {K\'{e}zsm\'{a}rki}\ \emph {et~al.}(2015)\citenamefont
  {K\'{e}zsm\'{a}rki}, \citenamefont {Bord\'{a}cs}, \citenamefont {Milde},
  \citenamefont {Neuber}, \citenamefont {Eng}, \citenamefont {White},
  \citenamefont {R{\o}nnow}, \citenamefont {Dewhurst}, \citenamefont
  {Mochizuki}, \citenamefont {Yanai}, \citenamefont {Nakamura}, \citenamefont
  {Ehlers}, \citenamefont {Tsurkan},\ and\ \citenamefont
  {Loidl}}]{Kezsmarki:2015bw}%
  \BibitemOpen
  \bibfield  {author} {\bibinfo {author} {\bibfnamefont {I.}~\bibnamefont
  {K\'{e}zsm\'{a}rki}}, \bibinfo {author} {\bibfnamefont {S.}~\bibnamefont
  {Bord\'{a}cs}}, \bibinfo {author} {\bibfnamefont {P.}~\bibnamefont {Milde}},
  \bibinfo {author} {\bibfnamefont {E.}~\bibnamefont {Neuber}}, \bibinfo
  {author} {\bibfnamefont {L.~M.}\ \bibnamefont {Eng}}, \bibinfo {author}
  {\bibfnamefont {J.~S.}\ \bibnamefont {White}}, \bibinfo {author}
  {\bibfnamefont {H.~M.}\ \bibnamefont {R{\o}nnow}}, \bibinfo {author}
  {\bibfnamefont {C.~D.}\ \bibnamefont {Dewhurst}}, \bibinfo {author}
  {\bibfnamefont {M.}~\bibnamefont {Mochizuki}}, \bibinfo {author}
  {\bibfnamefont {K.}~\bibnamefont {Yanai}}, \bibinfo {author} {\bibfnamefont
  {H.}~\bibnamefont {Nakamura}}, \bibinfo {author} {\bibfnamefont
  {D.}~\bibnamefont {Ehlers}}, \bibinfo {author} {\bibfnamefont
  {V.}~\bibnamefont {Tsurkan}},\ and\ \bibinfo {author} {\bibfnamefont
  {A.}~\bibnamefont {Loidl}},\ }\bibfield  {title} {\bibinfo {title}
  {{N\'{e}el-type skyrmion lattice with confined orientation in the polar
  magnetic semiconductor GaV$_4$S$_8$}},\ }\href
  {https://doi.org/10.1038/nmat4402} {\bibfield  {journal} {\bibinfo  {journal}
  {Nat. Mater.}\ }\textbf {\bibinfo {volume} {14}},\ \bibinfo {pages} {1116}
  (\bibinfo {year} {2015})}\BibitemShut {NoStop}%
\bibitem [{\citenamefont {Kurumaji}\ \emph {et~al.}(2019)\citenamefont
  {Kurumaji}, \citenamefont {Nakajima}, \citenamefont {Hirschberger},
  \citenamefont {Kikkawa}, \citenamefont {Yamasaki}, \citenamefont {Sagayama},
  \citenamefont {Nakao}, \citenamefont {Taguchi}, \citenamefont {Arima},\ and\
  \citenamefont {Tokura}}]{Kurumaji2019}%
  \BibitemOpen
  \bibfield  {author} {\bibinfo {author} {\bibfnamefont {T.}~\bibnamefont
  {Kurumaji}}, \bibinfo {author} {\bibfnamefont {T.}~\bibnamefont {Nakajima}},
  \bibinfo {author} {\bibfnamefont {M.}~\bibnamefont {Hirschberger}}, \bibinfo
  {author} {\bibfnamefont {A.}~\bibnamefont {Kikkawa}}, \bibinfo {author}
  {\bibfnamefont {Y.}~\bibnamefont {Yamasaki}}, \bibinfo {author}
  {\bibfnamefont {H.}~\bibnamefont {Sagayama}}, \bibinfo {author}
  {\bibfnamefont {H.}~\bibnamefont {Nakao}}, \bibinfo {author} {\bibfnamefont
  {Y.}~\bibnamefont {Taguchi}}, \bibinfo {author} {\bibfnamefont
  {T.}~\bibnamefont {Arima}},\ and\ \bibinfo {author} {\bibfnamefont
  {Y.}~\bibnamefont {Tokura}},\ }\bibfield  {title} {\bibinfo {title}
  {{Skyrmion lattice with a giant topological Hall effect in a frustrated
  triangular-lattice magnet}},\ }\href
  {https://doi.org/10.1126/science.aau0968} {\bibfield  {journal} {\bibinfo
  {journal} {Science}\ }\textbf {\bibinfo {volume} {365}},\ \bibinfo {pages}
  {914} (\bibinfo {year} {2019})}\BibitemShut {NoStop}%
\bibitem [{\citenamefont {Khanh}\ \emph {et~al.}(2020)\citenamefont {Khanh},
  \citenamefont {Nakajima}, \citenamefont {Yu}, \citenamefont {Gao},
  \citenamefont {Shibata}, \citenamefont {Hirschberger}, \citenamefont
  {Yamasaki}, \citenamefont {Sagayama}, \citenamefont {Nakao}, \citenamefont
  {Peng},\ and\ \citenamefont {et~al.}}]{Khanh2020}%
  \BibitemOpen
  \bibfield  {author} {\bibinfo {author} {\bibfnamefont {N.~D.}\ \bibnamefont
  {Khanh}}, \bibinfo {author} {\bibfnamefont {T.}~\bibnamefont {Nakajima}},
  \bibinfo {author} {\bibfnamefont {X.}~\bibnamefont {Yu}}, \bibinfo {author}
  {\bibfnamefont {S.}~\bibnamefont {Gao}}, \bibinfo {author} {\bibfnamefont
  {K.}~\bibnamefont {Shibata}}, \bibinfo {author} {\bibfnamefont
  {M.}~\bibnamefont {Hirschberger}}, \bibinfo {author} {\bibfnamefont
  {Y.}~\bibnamefont {Yamasaki}}, \bibinfo {author} {\bibfnamefont
  {H.}~\bibnamefont {Sagayama}}, \bibinfo {author} {\bibfnamefont
  {H.}~\bibnamefont {Nakao}}, \bibinfo {author} {\bibfnamefont
  {L.}~\bibnamefont {Peng}},\ and\ \bibinfo {author} {\bibnamefont {et~al.}},\
  }\bibfield  {title} {\bibinfo {title} {{Nanometric Square Skyrmion lattice in
  a centrosymmetric tetragonal magnet}},\ }\href
  {https://doi.org/10.1038/s41565-020-0684-7} {\bibfield  {journal} {\bibinfo
  {journal} {Nat. Nanotechnol.}\ }\textbf {\bibinfo {volume} {15}},\ \bibinfo
  {pages} {444} (\bibinfo {year} {2020})}\BibitemShut {NoStop}%
\bibitem [{\citenamefont {Hirschberger}\ \emph {et~al.}(2020)\citenamefont
  {Hirschberger}, \citenamefont {Nakajima}, \citenamefont {Kriener},
  \citenamefont {Kurumaji}, \citenamefont {Spitz}, \citenamefont {Gao},
  \citenamefont {Kikkawa}, \citenamefont {Yamasaki}, \citenamefont {Sagayama},
  \citenamefont {Nakao}, \citenamefont {Ohira-Kawamura}, \citenamefont
  {Taguchi}, \citenamefont {Arima},\ and\ \citenamefont
  {Tokura}}]{Hirschberger2020}%
  \BibitemOpen
  \bibfield  {author} {\bibinfo {author} {\bibfnamefont {M.}~\bibnamefont
  {Hirschberger}}, \bibinfo {author} {\bibfnamefont {T.}~\bibnamefont
  {Nakajima}}, \bibinfo {author} {\bibfnamefont {M.}~\bibnamefont {Kriener}},
  \bibinfo {author} {\bibfnamefont {T.}~\bibnamefont {Kurumaji}}, \bibinfo
  {author} {\bibfnamefont {L.}~\bibnamefont {Spitz}}, \bibinfo {author}
  {\bibfnamefont {S.}~\bibnamefont {Gao}}, \bibinfo {author} {\bibfnamefont
  {A.}~\bibnamefont {Kikkawa}}, \bibinfo {author} {\bibfnamefont
  {Y.}~\bibnamefont {Yamasaki}}, \bibinfo {author} {\bibfnamefont
  {H.}~\bibnamefont {Sagayama}}, \bibinfo {author} {\bibfnamefont
  {H.}~\bibnamefont {Nakao}}, \bibinfo {author} {\bibfnamefont
  {S.}~\bibnamefont {Ohira-Kawamura}}, \bibinfo {author} {\bibfnamefont
  {Y.}~\bibnamefont {Taguchi}}, \bibinfo {author} {\bibfnamefont
  {T.}~\bibnamefont {Arima}},\ and\ \bibinfo {author} {\bibfnamefont
  {Y.}~\bibnamefont {Tokura}},\ }\bibfield  {title} {\bibinfo {title}
  {{High-field depinned phase and planar Hall effect in the skyrmion host
  Gd\textsubscript{2}PdSi\textsubscript{3}}},\ }\href
  {https://doi.org/10.1103/PhysRevB.101.220401} {\bibfield  {journal} {\bibinfo
   {journal} {Phys. Rev. B}\ }\textbf {\bibinfo {volume} {101}},\ \bibinfo
  {pages} {220401} (\bibinfo {year} {2020})}\BibitemShut {NoStop}%
\bibitem [{\citenamefont {Casas}\ \emph {et~al.}(2023)\citenamefont {Casas},
  \citenamefont {Li}, \citenamefont {Moon}, \citenamefont {Xin}, \citenamefont
  {McKeever}, \citenamefont {Macy}, \citenamefont {Petford-Long}, \citenamefont
  {Phatak}, \citenamefont {Santos}, \citenamefont {Choi},\ and\ \citenamefont
  {Balicas}}]{Casas2023}%
  \BibitemOpen
  \bibfield  {author} {\bibinfo {author} {\bibfnamefont {B.~W.}\ \bibnamefont
  {Casas}}, \bibinfo {author} {\bibfnamefont {Y.}~\bibnamefont {Li}}, \bibinfo
  {author} {\bibfnamefont {A.}~\bibnamefont {Moon}}, \bibinfo {author}
  {\bibfnamefont {Y.}~\bibnamefont {Xin}}, \bibinfo {author} {\bibfnamefont
  {C.}~\bibnamefont {McKeever}}, \bibinfo {author} {\bibfnamefont
  {J.}~\bibnamefont {Macy}}, \bibinfo {author} {\bibfnamefont {A.~K.}\
  \bibnamefont {Petford-Long}}, \bibinfo {author} {\bibfnamefont {C.~M.}\
  \bibnamefont {Phatak}}, \bibinfo {author} {\bibfnamefont {E.~J.~G.}\
  \bibnamefont {Santos}}, \bibinfo {author} {\bibfnamefont {E.~S.}\
  \bibnamefont {Choi}},\ and\ \bibinfo {author} {\bibfnamefont
  {L.}~\bibnamefont {Balicas}},\ }\bibfield  {title} {\bibinfo {title}
  {{Coexistence of merons with skyrmions in the centrosymmetric van der Waals
  ferromagnet Fe\textsubscript{5}GeTe\textsubscript{2}}},\ }\href
  {https://doi.org/10.1002/adma.202212087} {\bibfield  {journal} {\bibinfo
  {journal} {Adv. Mater.}\ }\textbf {\bibinfo {volume} {35}},\ \bibinfo {pages}
  {2212087} (\bibinfo {year} {2023})}\BibitemShut {NoStop}%
\bibitem [{\citenamefont {Jonietz}\ \emph {et~al.}(2010)\citenamefont
  {Jonietz}, \citenamefont {Mühlbauer}, \citenamefont {Pfleiderer},
  \citenamefont {Neubauer}, \citenamefont {Münzer}, \citenamefont {Bauer},
  \citenamefont {Adams}, \citenamefont {Georgii}, \citenamefont {Böni},
  \citenamefont {Duine}, \citenamefont {Everschor}, \citenamefont {Garst},\
  and\ \citenamefont {Rosch}}]{Jonietz2010}%
  \BibitemOpen
  \bibfield  {author} {\bibinfo {author} {\bibfnamefont {F.}~\bibnamefont
  {Jonietz}}, \bibinfo {author} {\bibfnamefont {S.}~\bibnamefont {Mühlbauer}},
  \bibinfo {author} {\bibfnamefont {C.}~\bibnamefont {Pfleiderer}}, \bibinfo
  {author} {\bibfnamefont {A.}~\bibnamefont {Neubauer}}, \bibinfo {author}
  {\bibfnamefont {W.}~\bibnamefont {Münzer}}, \bibinfo {author} {\bibfnamefont
  {A.}~\bibnamefont {Bauer}}, \bibinfo {author} {\bibfnamefont
  {T.}~\bibnamefont {Adams}}, \bibinfo {author} {\bibfnamefont
  {R.}~\bibnamefont {Georgii}}, \bibinfo {author} {\bibfnamefont
  {P.}~\bibnamefont {Böni}}, \bibinfo {author} {\bibfnamefont {R.~A.}\
  \bibnamefont {Duine}}, \bibinfo {author} {\bibfnamefont {K.}~\bibnamefont
  {Everschor}}, \bibinfo {author} {\bibfnamefont {M.}~\bibnamefont {Garst}},\
  and\ \bibinfo {author} {\bibfnamefont {A.}~\bibnamefont {Rosch}},\ }\bibfield
   {title} {\bibinfo {title} {{Spin Transfer Torques in MnSi at Ultralow
  Current Densities}},\ }\href {https://doi.org/10.1126/science.1195709}
  {\bibfield  {journal} {\bibinfo  {journal} {Science}\ }\textbf {\bibinfo
  {volume} {330}},\ \bibinfo {pages} {1648} (\bibinfo {year}
  {2010})}\BibitemShut {NoStop}%
\bibitem [{\citenamefont {Okuyama}\ \emph {et~al.}(2019)\citenamefont
  {Okuyama}, \citenamefont {Bleuel}, \citenamefont {White}, \citenamefont {Ye},
  \citenamefont {Krzywon}, \citenamefont {Nagy}, \citenamefont {Im},
  \citenamefont {Zivkovic}, \citenamefont {Bartkowiak}, \citenamefont
  {R{\o}nnow}, \citenamefont {Hoshino}, \citenamefont {Iwasaki}, \citenamefont
  {Nagaosa}, \citenamefont {Kikkawa}, \citenamefont {Taguchi}, \citenamefont
  {Tokura}, \citenamefont {Higashi}, \citenamefont {Reim}, \citenamefont
  {Nambu},\ and\ \citenamefont {Sato}}]{Okuyama2019}%
  \BibitemOpen
  \bibfield  {author} {\bibinfo {author} {\bibfnamefont {D.}~\bibnamefont
  {Okuyama}}, \bibinfo {author} {\bibfnamefont {M.}~\bibnamefont {Bleuel}},
  \bibinfo {author} {\bibfnamefont {J.}~\bibnamefont {White}}, \bibinfo
  {author} {\bibfnamefont {Q.}~\bibnamefont {Ye}}, \bibinfo {author}
  {\bibfnamefont {J.}~\bibnamefont {Krzywon}}, \bibinfo {author} {\bibfnamefont
  {G.}~\bibnamefont {Nagy}}, \bibinfo {author} {\bibfnamefont {Z.}~\bibnamefont
  {Im}}, \bibinfo {author} {\bibfnamefont {I.}~\bibnamefont {Zivkovic}},
  \bibinfo {author} {\bibfnamefont {M.}~\bibnamefont {Bartkowiak}}, \bibinfo
  {author} {\bibfnamefont {H.}~\bibnamefont {R{\o}nnow}}, \bibinfo {author}
  {\bibfnamefont {S.}~\bibnamefont {Hoshino}}, \bibinfo {author} {\bibfnamefont
  {J.}~\bibnamefont {Iwasaki}}, \bibinfo {author} {\bibfnamefont
  {N.}~\bibnamefont {Nagaosa}}, \bibinfo {author} {\bibfnamefont
  {A.}~\bibnamefont {Kikkawa}}, \bibinfo {author} {\bibfnamefont
  {Y.}~\bibnamefont {Taguchi}}, \bibinfo {author} {\bibfnamefont
  {Y.}~\bibnamefont {Tokura}}, \bibinfo {author} {\bibfnamefont
  {D.}~\bibnamefont {Higashi}}, \bibinfo {author} {\bibfnamefont
  {J.}~\bibnamefont {Reim}}, \bibinfo {author} {\bibfnamefont {Y.}~\bibnamefont
  {Nambu}},\ and\ \bibinfo {author} {\bibfnamefont {T.}~\bibnamefont {Sato}},\
  }\bibfield  {title} {\bibinfo {title} {{Deformation of the moving magnetic
  skyrmion lattice in MnSi under electric current flow}},\ }\href
  {https://doi.org/10.1038/s42005-019-0175-z} {\bibfield  {journal} {\bibinfo
  {journal} {Commun. Phys.}\ }\textbf {\bibinfo {volume} {2}},\ \bibinfo
  {pages} {79} (\bibinfo {year} {2019})}\BibitemShut {NoStop}%
\bibitem [{\citenamefont {Back}\ \emph {et~al.}(2020)\citenamefont {Back},
  \citenamefont {Cros}, \citenamefont {Ebert}, \citenamefont {Everschor-Sitte},
  \citenamefont {Fert}, \citenamefont {Garst}, \citenamefont {Ma},
  \citenamefont {Mankovsky}, \citenamefont {Monchesky}, \citenamefont
  {Mostovoy},\ and\ \citenamefont {et~al.}}]{Back2020}%
  \BibitemOpen
  \bibfield  {author} {\bibinfo {author} {\bibfnamefont {C.}~\bibnamefont
  {Back}}, \bibinfo {author} {\bibfnamefont {V.}~\bibnamefont {Cros}}, \bibinfo
  {author} {\bibfnamefont {H.}~\bibnamefont {Ebert}}, \bibinfo {author}
  {\bibfnamefont {K.}~\bibnamefont {Everschor-Sitte}}, \bibinfo {author}
  {\bibfnamefont {A.}~\bibnamefont {Fert}}, \bibinfo {author} {\bibfnamefont
  {M.}~\bibnamefont {Garst}}, \bibinfo {author} {\bibfnamefont
  {T.}~\bibnamefont {Ma}}, \bibinfo {author} {\bibfnamefont {S.}~\bibnamefont
  {Mankovsky}}, \bibinfo {author} {\bibfnamefont {T.~L.}\ \bibnamefont
  {Monchesky}}, \bibinfo {author} {\bibfnamefont {M.}~\bibnamefont
  {Mostovoy}},\ and\ \bibinfo {author} {\bibnamefont {et~al.}},\ }\bibfield
  {title} {\bibinfo {title} {{The 2020 skyrmionics roadmap}},\ }\href
  {https://doi.org/10.1088/1361-6463/ab8418} {\bibfield  {journal} {\bibinfo
  {journal} {J. Phys. D}\ }\textbf {\bibinfo {volume} {53}},\ \bibinfo {pages}
  {363001} (\bibinfo {year} {2020})}\BibitemShut {NoStop}%
\bibitem [{\citenamefont {Dupé}\ \emph {et~al.}(2016)\citenamefont {Dupé},
  \citenamefont {Bihlmayer}, \citenamefont {Böttcher}, \citenamefont
  {Blügel},\ and\ \citenamefont {Heinze}}]{Dupe2016}%
  \BibitemOpen
  \bibfield  {author} {\bibinfo {author} {\bibfnamefont {B.}~\bibnamefont
  {Dupé}}, \bibinfo {author} {\bibfnamefont {G.}~\bibnamefont {Bihlmayer}},
  \bibinfo {author} {\bibfnamefont {M.}~\bibnamefont {Böttcher}}, \bibinfo
  {author} {\bibfnamefont {S.}~\bibnamefont {Blügel}},\ and\ \bibinfo {author}
  {\bibfnamefont {S.}~\bibnamefont {Heinze}},\ }\bibfield  {title} {\bibinfo
  {title} {{Engineering skyrmions in transition-metal multilayers for
  Spintronics}},\ }\href {https://doi.org/10.1038/ncomms11779} {\bibfield
  {journal} {\bibinfo  {journal} {Nat. Commun.}\ }\textbf {\bibinfo {volume}
  {7}},\ \bibinfo {pages} {11779} (\bibinfo {year} {2016})}\BibitemShut
  {NoStop}%
\bibitem [{\citenamefont {Wiesendanger}(2016)}]{Wiesendanger2016}%
  \BibitemOpen
  \bibfield  {author} {\bibinfo {author} {\bibfnamefont {R.}~\bibnamefont
  {Wiesendanger}},\ }\bibfield  {title} {\bibinfo {title} {{Nanoscale magnetic
  skyrmions in Metallic Films and Multilayers: A new twist for Spintronics}},\
  }\href {https://doi.org/10.1038/natrevmats.2016.44} {\bibfield  {journal}
  {\bibinfo  {journal} {Nat. Rev. Mater.}\ }\textbf {\bibinfo {volume} {1}},\
  \bibinfo {pages} {16044} (\bibinfo {year} {2016})}\BibitemShut {NoStop}%
\bibitem [{\citenamefont {Fert}\ \emph {et~al.}(2017)\citenamefont {Fert},
  \citenamefont {Reyren},\ and\ \citenamefont {Cros}}]{Fert2017}%
  \BibitemOpen
  \bibfield  {author} {\bibinfo {author} {\bibfnamefont {A.}~\bibnamefont
  {Fert}}, \bibinfo {author} {\bibfnamefont {N.}~\bibnamefont {Reyren}},\ and\
  \bibinfo {author} {\bibfnamefont {V.}~\bibnamefont {Cros}},\ }\bibfield
  {title} {\bibinfo {title} {{Magnetic skyrmions: Advances in physics and
  potential applications}},\ }\href
  {https://doi.org/10.1038/natrevmats.2017.31} {\bibfield  {journal} {\bibinfo
  {journal} {Nat. Rev. Mater.}\ }\textbf {\bibinfo {volume} {2}},\ \bibinfo
  {pages} {17031} (\bibinfo {year} {2017})}\BibitemShut {NoStop}%
\bibitem [{\citenamefont {Gibertini}\ \emph {et~al.}(2019)\citenamefont
  {Gibertini}, \citenamefont {Koperski}, \citenamefont {Morpurgo},\ and\
  \citenamefont {Novoselov}}]{Gibertini2019}%
  \BibitemOpen
  \bibfield  {author} {\bibinfo {author} {\bibfnamefont {M.}~\bibnamefont
  {Gibertini}}, \bibinfo {author} {\bibfnamefont {M.}~\bibnamefont {Koperski}},
  \bibinfo {author} {\bibfnamefont {A.~F.}\ \bibnamefont {Morpurgo}},\ and\
  \bibinfo {author} {\bibfnamefont {K.~S.}\ \bibnamefont {Novoselov}},\
  }\bibfield  {title} {\bibinfo {title} {{Magnetic 2D materials and
  heterostructures}},\ }\href {https://doi.org/10.1038/s41565-019-0438-6}
  {\bibfield  {journal} {\bibinfo  {journal} {Nature Nanotechnol.}\ }\textbf
  {\bibinfo {volume} {14}},\ \bibinfo {pages} {408} (\bibinfo {year}
  {2019})}\BibitemShut {NoStop}%
\bibitem [{\citenamefont {Fert}\ \emph {et~al.}(2013)\citenamefont {Fert},
  \citenamefont {Cros},\ and\ \citenamefont {Sampaio}}]{Fert2013}%
  \BibitemOpen
  \bibfield  {author} {\bibinfo {author} {\bibfnamefont {A.}~\bibnamefont
  {Fert}}, \bibinfo {author} {\bibfnamefont {V.}~\bibnamefont {Cros}},\ and\
  \bibinfo {author} {\bibfnamefont {J.}~\bibnamefont {Sampaio}},\ }\bibfield
  {title} {\bibinfo {title} {{Skyrmions on the track}},\ }\href
  {https://doi.org/10.1038/nnano.2013.29} {\bibfield  {journal} {\bibinfo
  {journal} {Nature Nanotech}\ }\textbf {\bibinfo {volume} {8}},\ \bibinfo
  {pages} {152} (\bibinfo {year} {2013})}\BibitemShut {NoStop}%
\bibitem [{\citenamefont {Schwarze}\ \emph {et~al.}(2015)\citenamefont
  {Schwarze}, \citenamefont {Waizner}, \citenamefont {Garst}, \citenamefont
  {Bauer}, \citenamefont {Stasinopoulos}, \citenamefont {Berger}, \citenamefont
  {Pfleiderer},\ and\ \citenamefont {Grundler}}]{Schwarze2015}%
  \BibitemOpen
  \bibfield  {author} {\bibinfo {author} {\bibfnamefont {T.}~\bibnamefont
  {Schwarze}}, \bibinfo {author} {\bibfnamefont {J.}~\bibnamefont {Waizner}},
  \bibinfo {author} {\bibfnamefont {M.}~\bibnamefont {Garst}}, \bibinfo
  {author} {\bibfnamefont {A.}~\bibnamefont {Bauer}}, \bibinfo {author}
  {\bibfnamefont {I.}~\bibnamefont {Stasinopoulos}}, \bibinfo {author}
  {\bibfnamefont {H.}~\bibnamefont {Berger}}, \bibinfo {author} {\bibfnamefont
  {C.}~\bibnamefont {Pfleiderer}},\ and\ \bibinfo {author} {\bibfnamefont
  {D.}~\bibnamefont {Grundler}},\ }\bibfield  {title} {\bibinfo {title}
  {{Universal helimagnon and skyrmion excitations in metallic, semiconducting
  and insulating chiral magnets}},\ }\href {https://doi.org/10.1038/nmat4223}
  {\bibfield  {journal} {\bibinfo  {journal} {Nat. Mater.}\ }\textbf {\bibinfo
  {volume} {14}},\ \bibinfo {pages} {478} (\bibinfo {year} {2015})}\BibitemShut
  {NoStop}%
\bibitem [{\citenamefont {M\"uhlbauer}\ \emph {et~al.}(2019)\citenamefont
  {M\"uhlbauer}, \citenamefont {Honecker}, \citenamefont {P\'erigo},
  \citenamefont {Bergner}, \citenamefont {Disch}, \citenamefont {Heinemann},
  \citenamefont {Erokhin}, \citenamefont {Berkov}, \citenamefont {Leighton},
  \citenamefont {Eskildsen},\ and\ \citenamefont {Michels}}]{Muhlbauer2019}%
  \BibitemOpen
  \bibfield  {author} {\bibinfo {author} {\bibfnamefont {S.}~\bibnamefont
  {M\"uhlbauer}}, \bibinfo {author} {\bibfnamefont {D.}~\bibnamefont
  {Honecker}}, \bibinfo {author} {\bibfnamefont {E.~A.}\ \bibnamefont
  {P\'erigo}}, \bibinfo {author} {\bibfnamefont {F.}~\bibnamefont {Bergner}},
  \bibinfo {author} {\bibfnamefont {S.}~\bibnamefont {Disch}}, \bibinfo
  {author} {\bibfnamefont {A.}~\bibnamefont {Heinemann}}, \bibinfo {author}
  {\bibfnamefont {S.}~\bibnamefont {Erokhin}}, \bibinfo {author} {\bibfnamefont
  {D.}~\bibnamefont {Berkov}}, \bibinfo {author} {\bibfnamefont
  {C.}~\bibnamefont {Leighton}}, \bibinfo {author} {\bibfnamefont {M.~R.}\
  \bibnamefont {Eskildsen}},\ and\ \bibinfo {author} {\bibfnamefont
  {A.}~\bibnamefont {Michels}},\ }\bibfield  {title} {\bibinfo {title}
  {{Magnetic small-angle neutron scattering}},\ }\href
  {https://doi.org/10.1103/RevModPhys.91.015004} {\bibfield  {journal}
  {\bibinfo  {journal} {Rev. Mod. Phys.}\ }\textbf {\bibinfo {volume} {91}},\
  \bibinfo {pages} {015004} (\bibinfo {year} {2019})}\BibitemShut {NoStop}%
\bibitem [{\citenamefont {White}\ \emph {et~al.}(2014)\citenamefont {White},
  \citenamefont {Pr\v{s}a}, \citenamefont {Huang}, \citenamefont {Omrani},
  \citenamefont {\v{Z}ivkovi\'{c}}, \citenamefont {Bartkowiak}, \citenamefont
  {Berger}, \citenamefont {Magrez}, \citenamefont {Gavilano}, \citenamefont
  {Nagy}, \citenamefont {Zang},\ and\ \citenamefont {R{\o}nnow}}]{White2014}%
  \BibitemOpen
  \bibfield  {author} {\bibinfo {author} {\bibfnamefont {J.~S.}\ \bibnamefont
  {White}}, \bibinfo {author} {\bibfnamefont {K.}~\bibnamefont {Pr\v{s}a}},
  \bibinfo {author} {\bibfnamefont {P.}~\bibnamefont {Huang}}, \bibinfo
  {author} {\bibfnamefont {A.~A.}\ \bibnamefont {Omrani}}, \bibinfo {author}
  {\bibfnamefont {I.}~\bibnamefont {\v{Z}ivkovi\'{c}}}, \bibinfo {author}
  {\bibfnamefont {M.}~\bibnamefont {Bartkowiak}}, \bibinfo {author}
  {\bibfnamefont {H.}~\bibnamefont {Berger}}, \bibinfo {author} {\bibfnamefont
  {A.}~\bibnamefont {Magrez}}, \bibinfo {author} {\bibfnamefont {J.~L.}\
  \bibnamefont {Gavilano}}, \bibinfo {author} {\bibfnamefont {G.}~\bibnamefont
  {Nagy}}, \bibinfo {author} {\bibfnamefont {J.}~\bibnamefont {Zang}},\ and\
  \bibinfo {author} {\bibfnamefont {H.~M.}\ \bibnamefont {R{\o}nnow}},\
  }\bibfield  {title} {\bibinfo {title} {{Electric-Field-Induced Skyrmion
  Distortion and Giant Lattice Rotation in the Magnetoelectric Insulator
  Cu$_2$OSeO$_3$}},\ }\href {https://doi.org/10.1103/PhysRevLett.113.107203}
  {\bibfield  {journal} {\bibinfo  {journal} {Phys. Rev. Lett.}\ }\textbf
  {\bibinfo {volume} {113}},\ \bibinfo {pages} {107203} (\bibinfo {year}
  {2014})}\BibitemShut {NoStop}%
\bibitem [{\citenamefont {Tengdin}\ \emph {et~al.}(2022)\citenamefont
  {Tengdin}, \citenamefont {Truc}, \citenamefont {Sapozhnik}, \citenamefont
  {Kong}, \citenamefont {del Ser}, \citenamefont {Gargiulo}, \citenamefont
  {Madan}, \citenamefont {Sch\"onenberger}, \citenamefont {Baral},
  \citenamefont {Che}, \citenamefont {Magrez}, \citenamefont {Grundler},
  \citenamefont {R{\o}nnow}, \citenamefont {Lagrange}, \citenamefont {Zang},
  \citenamefont {Rosch},\ and\ \citenamefont {Carbone}}]{Tengdin2022}%
  \BibitemOpen
  \bibfield  {author} {\bibinfo {author} {\bibfnamefont {P.}~\bibnamefont
  {Tengdin}}, \bibinfo {author} {\bibfnamefont {B.}~\bibnamefont {Truc}},
  \bibinfo {author} {\bibfnamefont {A.}~\bibnamefont {Sapozhnik}}, \bibinfo
  {author} {\bibfnamefont {L.}~\bibnamefont {Kong}}, \bibinfo {author}
  {\bibfnamefont {N.}~\bibnamefont {del Ser}}, \bibinfo {author} {\bibfnamefont
  {S.}~\bibnamefont {Gargiulo}}, \bibinfo {author} {\bibfnamefont
  {I.}~\bibnamefont {Madan}}, \bibinfo {author} {\bibfnamefont
  {T.}~\bibnamefont {Sch\"onenberger}}, \bibinfo {author} {\bibfnamefont
  {P.~R.}\ \bibnamefont {Baral}}, \bibinfo {author} {\bibfnamefont
  {P.}~\bibnamefont {Che}}, \bibinfo {author} {\bibfnamefont {A.}~\bibnamefont
  {Magrez}}, \bibinfo {author} {\bibfnamefont {D.}~\bibnamefont {Grundler}},
  \bibinfo {author} {\bibfnamefont {H.~M.}\ \bibnamefont {R{\o}nnow}}, \bibinfo
  {author} {\bibfnamefont {T.}~\bibnamefont {Lagrange}}, \bibinfo {author}
  {\bibfnamefont {J.}~\bibnamefont {Zang}}, \bibinfo {author} {\bibfnamefont
  {A.}~\bibnamefont {Rosch}},\ and\ \bibinfo {author} {\bibfnamefont
  {F.}~\bibnamefont {Carbone}},\ }\bibfield  {title} {\bibinfo {title}
  {{Imaging the Ultrafast Coherent Control of a Skyrmion Crystal}},\ }\href
  {https://doi.org/10.1103/PhysRevX.12.041030} {\bibfield  {journal} {\bibinfo
  {journal} {Phys. Rev. X}\ }\textbf {\bibinfo {volume} {12}},\ \bibinfo
  {pages} {041030} (\bibinfo {year} {2022})}\BibitemShut {NoStop}%
\bibitem [{\citenamefont {Everschor}\ \emph {et~al.}(2011)\citenamefont
  {Everschor}, \citenamefont {Garst}, \citenamefont {Duine},\ and\
  \citenamefont {Rosch}}]{Everschor2011}%
  \BibitemOpen
  \bibfield  {author} {\bibinfo {author} {\bibfnamefont {K.}~\bibnamefont
  {Everschor}}, \bibinfo {author} {\bibfnamefont {M.}~\bibnamefont {Garst}},
  \bibinfo {author} {\bibfnamefont {R.~A.}\ \bibnamefont {Duine}},\ and\
  \bibinfo {author} {\bibfnamefont {A.}~\bibnamefont {Rosch}},\ }\bibfield
  {title} {\bibinfo {title} {{Current-induced rotational torques in the
  Skyrmion lattice phase of chiral magnets}},\ }\href
  {https://doi.org/10.1103/physrevb.84.064401} {\bibfield  {journal} {\bibinfo
  {journal} {Phys. Rev. B}\ }\textbf {\bibinfo {volume} {84}},\ \bibinfo
  {pages} {064401} (\bibinfo {year} {2011})}\BibitemShut {NoStop}%
\bibitem [{\citenamefont {Everschor}\ \emph {et~al.}(2012)\citenamefont
  {Everschor}, \citenamefont {Garst}, \citenamefont {Binz}, \citenamefont
  {Jonietz}, \citenamefont {Mühlbauer}, \citenamefont {Pfleiderer},\ and\
  \citenamefont {Rosch}}]{Everschor2012}%
  \BibitemOpen
  \bibfield  {author} {\bibinfo {author} {\bibfnamefont {K.}~\bibnamefont
  {Everschor}}, \bibinfo {author} {\bibfnamefont {M.}~\bibnamefont {Garst}},
  \bibinfo {author} {\bibfnamefont {B.}~\bibnamefont {Binz}}, \bibinfo {author}
  {\bibfnamefont {F.}~\bibnamefont {Jonietz}}, \bibinfo {author} {\bibfnamefont
  {S.}~\bibnamefont {Mühlbauer}}, \bibinfo {author} {\bibfnamefont
  {C.}~\bibnamefont {Pfleiderer}},\ and\ \bibinfo {author} {\bibfnamefont
  {A.}~\bibnamefont {Rosch}},\ }\bibfield  {title} {\bibinfo {title} {{Rotating
  skyrmion lattices by spin torques and field or temperature gradients}},\
  }\href {https://doi.org/10.1103/physrevb.86.054432} {\bibfield  {journal}
  {\bibinfo  {journal} {Phys. Rev. B}\ }\textbf {\bibinfo {volume} {86}},\
  \bibinfo {pages} {054432} (\bibinfo {year} {2012})}\BibitemShut {NoStop}%
\bibitem [{\citenamefont {Mochizuki}\ \emph {et~al.}(2014)\citenamefont
  {Mochizuki}, \citenamefont {Yu}, \citenamefont {Seki}, \citenamefont
  {Kanazawa}, \citenamefont {Koshibae}, \citenamefont {Zang}, \citenamefont
  {Mostovoy}, \citenamefont {Tokura},\ and\ \citenamefont
  {Nagaosa}}]{Mochizuki:2014fja}%
  \BibitemOpen
  \bibfield  {author} {\bibinfo {author} {\bibfnamefont {M.}~\bibnamefont
  {Mochizuki}}, \bibinfo {author} {\bibfnamefont {X.~Z.}\ \bibnamefont {Yu}},
  \bibinfo {author} {\bibfnamefont {S.}~\bibnamefont {Seki}}, \bibinfo {author}
  {\bibfnamefont {N.}~\bibnamefont {Kanazawa}}, \bibinfo {author}
  {\bibfnamefont {W.}~\bibnamefont {Koshibae}}, \bibinfo {author}
  {\bibfnamefont {J.}~\bibnamefont {Zang}}, \bibinfo {author} {\bibfnamefont
  {M.}~\bibnamefont {Mostovoy}}, \bibinfo {author} {\bibfnamefont
  {Y.}~\bibnamefont {Tokura}},\ and\ \bibinfo {author} {\bibfnamefont
  {N.}~\bibnamefont {Nagaosa}},\ }\bibfield  {title} {\bibinfo {title}
  {{Thermally driven ratchet motion of a skyrmion microcrystal and topological
  magnon Hall effect}},\ }\href {https://doi.org/10.1038/nmat3862} {\bibfield
  {journal} {\bibinfo  {journal} {Nat. Mater.}\ }\textbf {\bibinfo {volume}
  {13}},\ \bibinfo {pages} {241 } (\bibinfo {year} {2014})}\BibitemShut
  {NoStop}%
\bibitem [{\citenamefont {P\"ollath}\ \emph {et~al.}(2017)\citenamefont
  {P\"ollath}, \citenamefont {Wild}, \citenamefont {Heinen}, \citenamefont
  {Meier}, \citenamefont {Kronseder}, \citenamefont {Tutsch}, \citenamefont
  {Bauer}, \citenamefont {Berger}, \citenamefont {Pfleiderer}, \citenamefont
  {Zweck}, \citenamefont {Rosch},\ and\ \citenamefont {Back}}]{Pollath2017}%
  \BibitemOpen
  \bibfield  {author} {\bibinfo {author} {\bibfnamefont {S.}~\bibnamefont
  {P\"ollath}}, \bibinfo {author} {\bibfnamefont {J.}~\bibnamefont {Wild}},
  \bibinfo {author} {\bibfnamefont {L.}~\bibnamefont {Heinen}}, \bibinfo
  {author} {\bibfnamefont {T.~N.~G.}\ \bibnamefont {Meier}}, \bibinfo {author}
  {\bibfnamefont {M.}~\bibnamefont {Kronseder}}, \bibinfo {author}
  {\bibfnamefont {L.}~\bibnamefont {Tutsch}}, \bibinfo {author} {\bibfnamefont
  {A.}~\bibnamefont {Bauer}}, \bibinfo {author} {\bibfnamefont
  {H.}~\bibnamefont {Berger}}, \bibinfo {author} {\bibfnamefont
  {C.}~\bibnamefont {Pfleiderer}}, \bibinfo {author} {\bibfnamefont
  {J.}~\bibnamefont {Zweck}}, \bibinfo {author} {\bibfnamefont
  {A.}~\bibnamefont {Rosch}},\ and\ \bibinfo {author} {\bibfnamefont {C.~H.}\
  \bibnamefont {Back}},\ }\bibfield  {title} {\bibinfo {title} {{Dynamical
  Defects in Rotating Magnetic Skyrmion Lattices}},\ }\href
  {https://doi.org/10.1103/PhysRevLett.118.207205} {\bibfield  {journal}
  {\bibinfo  {journal} {Phys. Rev. Lett.}\ }\textbf {\bibinfo {volume} {118}},\
  \bibinfo {pages} {207205} (\bibinfo {year} {2017})}\BibitemShut {NoStop}%
\bibitem [{\citenamefont {Yu}\ \emph {et~al.}(2021)\citenamefont {Yu},
  \citenamefont {Kagawa}, \citenamefont {Seki}, \citenamefont {Kubota},
  \citenamefont {Masell}, \citenamefont {Yasin}, \citenamefont {Nakajima},
  \citenamefont {Nakamura}, \citenamefont {Kawasaki}, \citenamefont {Nagaosa},\
  and\ \citenamefont {Tokura}}]{Yu2021}%
  \BibitemOpen
  \bibfield  {author} {\bibinfo {author} {\bibfnamefont {X.}~\bibnamefont
  {Yu}}, \bibinfo {author} {\bibfnamefont {F.}~\bibnamefont {Kagawa}}, \bibinfo
  {author} {\bibfnamefont {S.}~\bibnamefont {Seki}}, \bibinfo {author}
  {\bibfnamefont {M.}~\bibnamefont {Kubota}}, \bibinfo {author} {\bibfnamefont
  {J.}~\bibnamefont {Masell}}, \bibinfo {author} {\bibfnamefont {F.~S.}\
  \bibnamefont {Yasin}}, \bibinfo {author} {\bibfnamefont {K.}~\bibnamefont
  {Nakajima}}, \bibinfo {author} {\bibfnamefont {M.}~\bibnamefont {Nakamura}},
  \bibinfo {author} {\bibfnamefont {M.}~\bibnamefont {Kawasaki}}, \bibinfo
  {author} {\bibfnamefont {N.}~\bibnamefont {Nagaosa}},\ and\ \bibinfo {author}
  {\bibfnamefont {Y.}~\bibnamefont {Tokura}},\ }\bibfield  {title} {\bibinfo
  {title} {{Real-space observations of 60-nm skyrmion dynamics in an insulating
  magnet under low heat flow}},\ }\href
  {https://doi.org/10.1038/s41467-021-25291-2} {\bibfield  {journal} {\bibinfo
  {journal} {Nat. Commun.}\ }\textbf {\bibinfo {volume} {12}},\ \bibinfo
  {pages} {5079} (\bibinfo {year} {2021})}\BibitemShut {NoStop}%
\bibitem [{\citenamefont {Yu}\ \emph {et~al.}(2010{\natexlab{b}})\citenamefont
  {Yu}, \citenamefont {Onose}, \citenamefont {Kanazawa}, \citenamefont {Park},
  \citenamefont {Han}, \citenamefont {Matsui}, \citenamefont {Nagaosa},\ and\
  \citenamefont {Tokura}}]{Yu2010}%
  \BibitemOpen
  \bibfield  {author} {\bibinfo {author} {\bibfnamefont {X.~Z.}\ \bibnamefont
  {Yu}}, \bibinfo {author} {\bibfnamefont {Y.}~\bibnamefont {Onose}}, \bibinfo
  {author} {\bibfnamefont {N.}~\bibnamefont {Kanazawa}}, \bibinfo {author}
  {\bibfnamefont {J.~H.}\ \bibnamefont {Park}}, \bibinfo {author}
  {\bibfnamefont {J.~H.}\ \bibnamefont {Han}}, \bibinfo {author} {\bibfnamefont
  {Y.}~\bibnamefont {Matsui}}, \bibinfo {author} {\bibfnamefont
  {N.}~\bibnamefont {Nagaosa}},\ and\ \bibinfo {author} {\bibfnamefont
  {Y.}~\bibnamefont {Tokura}},\ }\bibfield  {title} {\bibinfo {title}
  {{Real-space observation of a two-dimensional Skyrmion Crystal}},\ }\href
  {https://doi.org/10.1038/nature09124} {\bibfield  {journal} {\bibinfo
  {journal} {Nature}\ }\textbf {\bibinfo {volume} {465}},\ \bibinfo {pages}
  {901} (\bibinfo {year} {2010}{\natexlab{b}})}\BibitemShut {NoStop}%
\bibitem [{\citenamefont {Milde}\ \emph {et~al.}(2013)\citenamefont {Milde},
  \citenamefont {Köhler}, \citenamefont {Seidel}, \citenamefont {Eng},
  \citenamefont {Bauer}, \citenamefont {Chacon}, \citenamefont {Kindervater},
  \citenamefont {Mühlbauer}, \citenamefont {Pfleiderer}, \citenamefont
  {Buhrandt}, \citenamefont {Schütte},\ and\ \citenamefont
  {Rosch}}]{Milde2013}%
  \BibitemOpen
  \bibfield  {author} {\bibinfo {author} {\bibfnamefont {P.}~\bibnamefont
  {Milde}}, \bibinfo {author} {\bibfnamefont {D.}~\bibnamefont {Köhler}},
  \bibinfo {author} {\bibfnamefont {J.}~\bibnamefont {Seidel}}, \bibinfo
  {author} {\bibfnamefont {L.~M.}\ \bibnamefont {Eng}}, \bibinfo {author}
  {\bibfnamefont {A.}~\bibnamefont {Bauer}}, \bibinfo {author} {\bibfnamefont
  {A.}~\bibnamefont {Chacon}}, \bibinfo {author} {\bibfnamefont
  {J.}~\bibnamefont {Kindervater}}, \bibinfo {author} {\bibfnamefont
  {S.}~\bibnamefont {Mühlbauer}}, \bibinfo {author} {\bibfnamefont
  {C.}~\bibnamefont {Pfleiderer}}, \bibinfo {author} {\bibfnamefont
  {S.}~\bibnamefont {Buhrandt}}, \bibinfo {author} {\bibfnamefont
  {C.}~\bibnamefont {Schütte}},\ and\ \bibinfo {author} {\bibfnamefont
  {A.}~\bibnamefont {Rosch}},\ }\bibfield  {title} {\bibinfo {title}
  {{Unwinding of a Skyrmion Lattice by Magnetic Monopoles}},\ }\href
  {https://doi.org/10.1126/science.1234657} {\bibfield  {journal} {\bibinfo
  {journal} {Science}\ }\textbf {\bibinfo {volume} {340}},\ \bibinfo {pages}
  {1076} (\bibinfo {year} {2013})}\BibitemShut {NoStop}%
\bibitem [{\citenamefont {Eskildsen}\ \emph {et~al.}(2021)\citenamefont
  {Eskildsen}, \citenamefont {Chalus}, \citenamefont {Cubitt}, \citenamefont
  {Janoschek}, \citenamefont {Leishman}, \citenamefont {Longbons},\ and\
  \citenamefont {White}}]{5-42-540}%
  \BibitemOpen
  \bibfield  {author} {\bibinfo {author} {\bibfnamefont {M.~R.}\ \bibnamefont
  {Eskildsen}}, \bibinfo {author} {\bibfnamefont {N.}~\bibnamefont {Chalus}},
  \bibinfo {author} {\bibfnamefont {R.}~\bibnamefont {Cubitt}}, \bibinfo
  {author} {\bibfnamefont {M.}~\bibnamefont {Janoschek}}, \bibinfo {author}
  {\bibfnamefont {A.~W.~D.}\ \bibnamefont {Leishman}}, \bibinfo {author}
  {\bibfnamefont {G.}~\bibnamefont {Longbons}},\ and\ \bibinfo {author}
  {\bibfnamefont {J.}~\bibnamefont {White}},\ }\href@noop {} {\bibinfo {title}
  {{Skyrmion manipulation using radial currents}}},\ \bibinfo {howpublished}
  {Institut Laue-Langevin (ILL) doi:10.5291/ILL-DATA.5-42-540} (\bibinfo {year}
  {2021})\BibitemShut {NoStop}%
\bibitem [{\citenamefont {Chalus}\ \emph {et~al.}(2023)\citenamefont {Chalus},
  \citenamefont {Cubitt}, \citenamefont {Eskildsen}, \citenamefont {Forgan},
  \citenamefont {Janoschek}, \citenamefont {Longbons}, \citenamefont
  {M.~Menezes}, \citenamefont {Milosevic}, \citenamefont {Powers},\ and\
  \citenamefont {White}}]{5-42-568}%
  \BibitemOpen
  \bibfield  {author} {\bibinfo {author} {\bibfnamefont {N.}~\bibnamefont
  {Chalus}}, \bibinfo {author} {\bibfnamefont {R.}~\bibnamefont {Cubitt}},
  \bibinfo {author} {\bibfnamefont {M.~R.}\ \bibnamefont {Eskildsen}}, \bibinfo
  {author} {\bibfnamefont {E.~M.}\ \bibnamefont {Forgan}}, \bibinfo {author}
  {\bibfnamefont {M.}~\bibnamefont {Janoschek}}, \bibinfo {author}
  {\bibfnamefont {G.~M.}\ \bibnamefont {Longbons}}, \bibinfo {author}
  {\bibfnamefont {R.}~\bibnamefont {M.~Menezes}}, \bibinfo {author}
  {\bibfnamefont {M.}~\bibnamefont {Milosevic}}, \bibinfo {author}
  {\bibfnamefont {N.}~\bibnamefont {Powers}},\ and\ \bibinfo {author}
  {\bibfnamefont {J.}~\bibnamefont {White}},\ }\href
  {https://doi.ill.fr/10.5291/ILL-DATA.5-42-568} {\bibinfo {title} {{Skyrmion
  Lattice Manipulation Using Electric and Thermal Currents in MnSi}}},\
  \bibinfo {howpublished} {Institut Laue-Langevin (ILL)
  doi:10.5291/ILL-DATA.5-42-568} (\bibinfo {year} {2023})\BibitemShut {NoStop}%
\bibitem [{\citenamefont {Leishman}\ \emph {et~al.}(2020)\citenamefont
  {Leishman}, \citenamefont {Menezes}, \citenamefont {Longbons}, \citenamefont
  {Bauer}, \citenamefont {Janoschek}, \citenamefont {Honecker}, \citenamefont
  {DeBeer-Schmitt}, \citenamefont {White}, \citenamefont {Sokolova},
  \citenamefont {V.},\ and\ \citenamefont {Eskildsen}}]{Leishman:2020tg}%
  \BibitemOpen
  \bibfield  {author} {\bibinfo {author} {\bibfnamefont {A.~W.~D.}\
  \bibnamefont {Leishman}}, \bibinfo {author} {\bibfnamefont {R.~M.}\
  \bibnamefont {Menezes}}, \bibinfo {author} {\bibfnamefont {G.}~\bibnamefont
  {Longbons}}, \bibinfo {author} {\bibfnamefont {E.~D.}\ \bibnamefont {Bauer}},
  \bibinfo {author} {\bibfnamefont {M.}~\bibnamefont {Janoschek}}, \bibinfo
  {author} {\bibfnamefont {D.}~\bibnamefont {Honecker}}, \bibinfo {author}
  {\bibfnamefont {L.}~\bibnamefont {DeBeer-Schmitt}}, \bibinfo {author}
  {\bibfnamefont {J.~S.}\ \bibnamefont {White}}, \bibinfo {author}
  {\bibfnamefont {A.}~\bibnamefont {Sokolova}}, \bibinfo {author}
  {\bibfnamefont {M.~M.}\ \bibnamefont {V.}},\ and\ \bibinfo {author}
  {\bibfnamefont {M.~R.}\ \bibnamefont {Eskildsen}},\ }\bibfield  {title}
  {\bibinfo {title} {{Topological energy barrier for skyrmion lattice formation
  in MnSi}},\ }\href {https://doi.org/10.1103/PhysRevB.102.104416} {\bibfield
  {journal} {\bibinfo  {journal} {Phys. Rev. B}\ }\textbf {\bibinfo {volume}
  {102}},\ \bibinfo {pages} {104416} (\bibinfo {year} {2020})}\BibitemShut
  {NoStop}%
\bibitem [{\citenamefont {Vansteenkiste}\ \emph {et~al.}(2014)\citenamefont
  {Vansteenkiste}, \citenamefont {Leliaert}, \citenamefont {Dvornik},
  \citenamefont {Helsen}, \citenamefont {Garcia-Sanchez},\ and\ \citenamefont
  {Van~Waeyenberge}}]{Vansteenkiste2014}%
  \BibitemOpen
  \bibfield  {author} {\bibinfo {author} {\bibfnamefont {A.}~\bibnamefont
  {Vansteenkiste}}, \bibinfo {author} {\bibfnamefont {J.}~\bibnamefont
  {Leliaert}}, \bibinfo {author} {\bibfnamefont {M.}~\bibnamefont {Dvornik}},
  \bibinfo {author} {\bibfnamefont {M.}~\bibnamefont {Helsen}}, \bibinfo
  {author} {\bibfnamefont {F.}~\bibnamefont {Garcia-Sanchez}},\ and\ \bibinfo
  {author} {\bibfnamefont {B.}~\bibnamefont {Van~Waeyenberge}},\ }\bibfield
  {title} {\bibinfo {title} {{The design and verification of MuMax3}},\ }\href
  {https://doi.org/10.1063/1.4899186} {\bibfield  {journal} {\bibinfo
  {journal} {AIP Adv.}\ }\textbf {\bibinfo {volume} {4}},\ \bibinfo {pages}
  {107133} (\bibinfo {year} {2014})}\BibitemShut {NoStop}%
\bibitem [{\citenamefont {Karhu}\ \emph {et~al.}(2012)\citenamefont {Karhu},
  \citenamefont {R{\"o}{\ss}ler}, \citenamefont {Bogdanov}, \citenamefont
  {Kahwaji}, \citenamefont {Kirby}, \citenamefont {Fritzsche}, \citenamefont
  {Robertson}, \citenamefont {Majkrzak},\ and\ \citenamefont
  {Monchesky}}]{Karhu2012}%
  \BibitemOpen
  \bibfield  {author} {\bibinfo {author} {\bibfnamefont {E.~A.}\ \bibnamefont
  {Karhu}}, \bibinfo {author} {\bibfnamefont {U.~K.}\ \bibnamefont
  {R{\"o}{\ss}ler}}, \bibinfo {author} {\bibfnamefont {A.~N.}\ \bibnamefont
  {Bogdanov}}, \bibinfo {author} {\bibfnamefont {S.}~\bibnamefont {Kahwaji}},
  \bibinfo {author} {\bibfnamefont {B.~J.}\ \bibnamefont {Kirby}}, \bibinfo
  {author} {\bibfnamefont {H.}~\bibnamefont {Fritzsche}}, \bibinfo {author}
  {\bibfnamefont {M.~D.}\ \bibnamefont {Robertson}}, \bibinfo {author}
  {\bibfnamefont {C.~F.}\ \bibnamefont {Majkrzak}},\ and\ \bibinfo {author}
  {\bibfnamefont {T.~L.}\ \bibnamefont {Monchesky}},\ }\bibfield  {title}
  {\bibinfo {title} {{Chiral modulations and reorientation effects in MnSi thin
  films}},\ }\href {https://doi.org/10.1103/PhysRevB.85.094429} {\bibfield
  {journal} {\bibinfo  {journal} {Phys. Rev. B}\ }\textbf {\bibinfo {volume}
  {85}},\ \bibinfo {pages} {094429} (\bibinfo {year} {2012})}\BibitemShut
  {NoStop}%
\bibitem [{\citenamefont {Neubauer}\ \emph {et~al.}(2009)\citenamefont
  {Neubauer}, \citenamefont {Pfleiderer}, \citenamefont {Binz}, \citenamefont
  {Rosch}, \citenamefont {Ritz}, \citenamefont {Niklowitz},\ and\ \citenamefont
  {B{\"o}ni}}]{Neubauer2009}%
  \BibitemOpen
  \bibfield  {author} {\bibinfo {author} {\bibfnamefont {A.}~\bibnamefont
  {Neubauer}}, \bibinfo {author} {\bibfnamefont {C.}~\bibnamefont
  {Pfleiderer}}, \bibinfo {author} {\bibfnamefont {B.}~\bibnamefont {Binz}},
  \bibinfo {author} {\bibfnamefont {A.}~\bibnamefont {Rosch}}, \bibinfo
  {author} {\bibfnamefont {R.}~\bibnamefont {Ritz}}, \bibinfo {author}
  {\bibfnamefont {P.~G.}\ \bibnamefont {Niklowitz}},\ and\ \bibinfo {author}
  {\bibfnamefont {P.}~\bibnamefont {B{\"o}ni}},\ }\bibfield  {title} {\bibinfo
  {title} {{Topological Hall effect in the A phase of MnSi}},\ }\href
  {https://doi.org/10.1103/PhysRevLett.102.186602} {\bibfield  {journal}
  {\bibinfo  {journal} {Phys. Rev. Lett.}\ }\textbf {\bibinfo {volume} {102}},\
  \bibinfo {pages} {186602} (\bibinfo {year} {2009})}\BibitemShut {NoStop}%
\bibitem [{\citenamefont {Stishov}\ \emph {et~al.}(2008)\citenamefont
  {Stishov}, \citenamefont {Petrova}, \citenamefont {Khasanov}, \citenamefont
  {Kh~Panova}, \citenamefont {Shikov}, \citenamefont {Lashley}, \citenamefont
  {Wu},\ and\ \citenamefont {Lograsso}}]{Stishov2008}%
  \BibitemOpen
  \bibfield  {author} {\bibinfo {author} {\bibfnamefont {S.~M.}\ \bibnamefont
  {Stishov}}, \bibinfo {author} {\bibfnamefont {A.~E.}\ \bibnamefont
  {Petrova}}, \bibinfo {author} {\bibfnamefont {S.}~\bibnamefont {Khasanov}},
  \bibinfo {author} {\bibfnamefont {G.}~\bibnamefont {Kh~Panova}}, \bibinfo
  {author} {\bibfnamefont {A.~A.}\ \bibnamefont {Shikov}}, \bibinfo {author}
  {\bibfnamefont {J.~C.}\ \bibnamefont {Lashley}}, \bibinfo {author}
  {\bibfnamefont {D.}~\bibnamefont {Wu}},\ and\ \bibinfo {author}
  {\bibfnamefont {T.~A.}\ \bibnamefont {Lograsso}},\ }\bibfield  {title}
  {\bibinfo {title} {{Heat capacity and thermal expansion of the itinerant
  helimagnet MnSi}},\ }\href {https://doi.org/10.1088/0953-8984/20/23/235222}
  {\bibfield  {journal} {\bibinfo  {journal} {J. Phys. Condens. Matter}\
  }\textbf {\bibinfo {volume} {20}},\ \bibinfo {pages} {235222} (\bibinfo
  {year} {2008})}\BibitemShut {NoStop}%
\bibitem [{\citenamefont {Raimondo}\ \emph {et~al.}(2022)\citenamefont
  {Raimondo}, \citenamefont {Saugar}, \citenamefont {Barker}, \citenamefont
  {Rodrigues}, \citenamefont {Giordano}, \citenamefont {Carpentieri},
  \citenamefont {Jiang}, \citenamefont {Chubykalo-Fesenko}, \citenamefont
  {Tomasello},\ and\ \citenamefont {Finocchio}}]{raimondo2022temperature}%
  \BibitemOpen
  \bibfield  {author} {\bibinfo {author} {\bibfnamefont {E.}~\bibnamefont
  {Raimondo}}, \bibinfo {author} {\bibfnamefont {E.}~\bibnamefont {Saugar}},
  \bibinfo {author} {\bibfnamefont {J.}~\bibnamefont {Barker}}, \bibinfo
  {author} {\bibfnamefont {D.}~\bibnamefont {Rodrigues}}, \bibinfo {author}
  {\bibfnamefont {A.}~\bibnamefont {Giordano}}, \bibinfo {author}
  {\bibfnamefont {M.}~\bibnamefont {Carpentieri}}, \bibinfo {author}
  {\bibfnamefont {W.}~\bibnamefont {Jiang}}, \bibinfo {author} {\bibfnamefont
  {O.}~\bibnamefont {Chubykalo-Fesenko}}, \bibinfo {author} {\bibfnamefont
  {R.}~\bibnamefont {Tomasello}},\ and\ \bibinfo {author} {\bibfnamefont
  {G.}~\bibnamefont {Finocchio}},\ }\bibfield  {title} {\bibinfo {title}
  {Temperature-gradient-driven magnetic skyrmion motion},\ }\href@noop {}
  {\bibfield  {journal} {\bibinfo  {journal} {Physical Review Applied}\
  }\textbf {\bibinfo {volume} {18}},\ \bibinfo {pages} {024062} (\bibinfo
  {year} {2022})}\BibitemShut {NoStop}%
\bibitem [{\citenamefont {Menezes}\ and\ \citenamefont
  {Silva}(2017)}]{Menezes2017}%
  \BibitemOpen
  \bibfield  {author} {\bibinfo {author} {\bibfnamefont {R.~M.}\ \bibnamefont
  {Menezes}}\ and\ \bibinfo {author} {\bibfnamefont {C.~C. d.~S.}\ \bibnamefont
  {Silva}},\ }\bibfield  {title} {\bibinfo {title} {{Conformal vortex
  crystals}},\ }\href {https://doi.org/10.1038/s41598-017-12807-4} {\bibfield
  {journal} {\bibinfo  {journal} {Sci. Rep.}\ }\textbf {\bibinfo {volume}
  {7}},\ \bibinfo {pages} {12766} (\bibinfo {year} {2017})}\BibitemShut
  {NoStop}%
\bibitem [{\citenamefont {Braun}\ \emph {et~al.}(1996)\citenamefont {Braun},
  \citenamefont {Crabtree}, \citenamefont {Kaper}, \citenamefont {Koshelev},
  \citenamefont {Leaf}, \citenamefont {Levine},\ and\ \citenamefont
  {Vinokur}}]{Braun:1996tn}%
  \BibitemOpen
  \bibfield  {author} {\bibinfo {author} {\bibfnamefont {D.~W.}\ \bibnamefont
  {Braun}}, \bibinfo {author} {\bibfnamefont {G.~W.}\ \bibnamefont {Crabtree}},
  \bibinfo {author} {\bibfnamefont {H.~G.}\ \bibnamefont {Kaper}}, \bibinfo
  {author} {\bibfnamefont {A.~E.}\ \bibnamefont {Koshelev}}, \bibinfo {author}
  {\bibfnamefont {G.~K.}\ \bibnamefont {Leaf}}, \bibinfo {author}
  {\bibfnamefont {D.~M.}\ \bibnamefont {Levine}},\ and\ \bibinfo {author}
  {\bibfnamefont {V.~M.}\ \bibnamefont {Vinokur}},\ }\bibfield  {title}
  {\bibinfo {title} {{Structure of a Moving Vortex Lattice}},\ }\href
  {https://doi.org/10.1103/physrevlett.76.831} {\bibfield  {journal} {\bibinfo
  {journal} {Phys. Rev. Lett.}\ }\textbf {\bibinfo {volume} {76}},\ \bibinfo
  {pages} {831834} (\bibinfo {year} {1996})}\BibitemShut {NoStop}%
\bibitem [{\citenamefont {Zhang}\ \emph {et~al.}(2018)\citenamefont {Zhang},
  \citenamefont {Wang}, \citenamefont {Burn}, \citenamefont {Peng},
  \citenamefont {Berger}, \citenamefont {Bauer}, \citenamefont {Pfleiderer},
  \citenamefont {Laan},\ and\ \citenamefont {Hesjedal}}]{Zhang:2018bg}%
  \BibitemOpen
  \bibfield  {author} {\bibinfo {author} {\bibfnamefont {S.~L.}\ \bibnamefont
  {Zhang}}, \bibinfo {author} {\bibfnamefont {W.~W.}\ \bibnamefont {Wang}},
  \bibinfo {author} {\bibfnamefont {D.~M.}\ \bibnamefont {Burn}}, \bibinfo
  {author} {\bibfnamefont {H.}~\bibnamefont {Peng}}, \bibinfo {author}
  {\bibfnamefont {H.}~\bibnamefont {Berger}}, \bibinfo {author} {\bibfnamefont
  {A.}~\bibnamefont {Bauer}}, \bibinfo {author} {\bibfnamefont
  {C.}~\bibnamefont {Pfleiderer}}, \bibinfo {author} {\bibfnamefont {G.~v.~d.}\
  \bibnamefont {Laan}},\ and\ \bibinfo {author} {\bibfnamefont
  {T.}~\bibnamefont {Hesjedal}},\ }\bibfield  {title} {\bibinfo {title}
  {{Manipulation of skyrmion motion by magnetic field gradients}},\ }\href
  {https://doi.org/10.1038/s41467-018-04563-4} {\bibfield  {journal} {\bibinfo
  {journal} {Nat. Commun.}\ }\textbf {\bibinfo {volume} {9}},\ \bibinfo {pages}
  {2115} (\bibinfo {year} {2018})}\BibitemShut {NoStop}%
\bibitem [{\citenamefont {Jin}\ \emph {et~al.}(2024)\citenamefont {Jin},
  \citenamefont {Chen}, \citenamefont {Laan}, \citenamefont {Hesjedal},
  \citenamefont {Liu},\ and\ \citenamefont {Zhang}}]{Jin.2024ajk}%
  \BibitemOpen
  \bibfield  {author} {\bibinfo {author} {\bibfnamefont {H.}~\bibnamefont
  {Jin}}, \bibinfo {author} {\bibfnamefont {J.}~\bibnamefont {Chen}}, \bibinfo
  {author} {\bibfnamefont {G.~v.~d.}\ \bibnamefont {Laan}}, \bibinfo {author}
  {\bibfnamefont {T.}~\bibnamefont {Hesjedal}}, \bibinfo {author}
  {\bibfnamefont {Y.}~\bibnamefont {Liu}},\ and\ \bibinfo {author}
  {\bibfnamefont {S.}~\bibnamefont {Zhang}},\ }\bibfield  {title} {\bibinfo
  {title} {{Rolling Motion of Rigid Skyrmion Crystallites Induced by Chiral
  Lattice Torque}},\ }\href {https://doi.org/10.1021/acs.nanolett.4c03336}
  {\bibfield  {journal} {\bibinfo  {journal} {Nano Lett.}\ }\textbf {\bibinfo
  {volume} {24}},\ \bibinfo {pages} {12226} (\bibinfo {year}
  {2024})}\BibitemShut {NoStop}%
\bibitem [{\citenamefont {Yokouchi}\ \emph {et~al.}(2022)\citenamefont
  {Yokouchi}, \citenamefont {Sugimoto}, \citenamefont {Rana}, \citenamefont
  {Seki}, \citenamefont {Ogawa}, \citenamefont {Shiomi}, \citenamefont
  {Kasai},\ and\ \citenamefont {Otani}}]{yokouchi2022pattern}%
  \BibitemOpen
  \bibfield  {author} {\bibinfo {author} {\bibfnamefont {T.}~\bibnamefont
  {Yokouchi}}, \bibinfo {author} {\bibfnamefont {S.}~\bibnamefont {Sugimoto}},
  \bibinfo {author} {\bibfnamefont {B.}~\bibnamefont {Rana}}, \bibinfo {author}
  {\bibfnamefont {S.}~\bibnamefont {Seki}}, \bibinfo {author} {\bibfnamefont
  {N.}~\bibnamefont {Ogawa}}, \bibinfo {author} {\bibfnamefont
  {Y.}~\bibnamefont {Shiomi}}, \bibinfo {author} {\bibfnamefont
  {S.}~\bibnamefont {Kasai}},\ and\ \bibinfo {author} {\bibfnamefont
  {Y.}~\bibnamefont {Otani}},\ }\bibfield  {title} {\bibinfo {title} {Pattern
  recognition with neuromorphic computing using magnetic field--induced
  dynamics of skyrmions},\ }\href@noop {} {\bibfield  {journal} {\bibinfo
  {journal} {Science Advances}\ }\textbf {\bibinfo {volume} {8}},\ \bibinfo
  {pages} {eabq5652} (\bibinfo {year} {2022})}\BibitemShut {NoStop}%
\bibitem [{\citenamefont {Li}\ \emph {et~al.}(2021)\citenamefont {Li},
  \citenamefont {Kang}, \citenamefont {Zhang}, \citenamefont {Nie},
  \citenamefont {Zhou}, \citenamefont {Wang},\ and\ \citenamefont
  {Zhao}}]{li2021magnetic}%
  \BibitemOpen
  \bibfield  {author} {\bibinfo {author} {\bibfnamefont {S.}~\bibnamefont
  {Li}}, \bibinfo {author} {\bibfnamefont {W.}~\bibnamefont {Kang}}, \bibinfo
  {author} {\bibfnamefont {X.}~\bibnamefont {Zhang}}, \bibinfo {author}
  {\bibfnamefont {T.}~\bibnamefont {Nie}}, \bibinfo {author} {\bibfnamefont
  {Y.}~\bibnamefont {Zhou}}, \bibinfo {author} {\bibfnamefont {K.~L.}\
  \bibnamefont {Wang}},\ and\ \bibinfo {author} {\bibfnamefont
  {W.}~\bibnamefont {Zhao}},\ }\bibfield  {title} {\bibinfo {title} {Magnetic
  skyrmions for unconventional computing},\ }\href@noop {} {\bibfield
  {journal} {\bibinfo  {journal} {Materials Horizons}\ }\textbf {\bibinfo
  {volume} {8}},\ \bibinfo {pages} {854} (\bibinfo {year} {2021})}\BibitemShut
  {NoStop}%
\bibitem [{\citenamefont {{COMSOL AB}}()}]{comsol}%
  \BibitemOpen
  \bibfield  {author} {\bibinfo {author} {\bibnamefont {{COMSOL AB}}},\ }\href
  {https://www.comsol.com} {\bibinfo {title} {{COMSOL
  Multiphysics\textsuperscript{\textregistered} v. 6.1}}},\ \bibinfo {note}
  {{Stockholm, Sweden}}\BibitemShut {NoStop}%
\bibitem [{\citenamefont {Cheng}\ \emph {et~al.}(2010)\citenamefont {Cheng},
  \citenamefont {Zhou}, \citenamefont {Zhou}, \citenamefont {Goodenough},\ and\
  \citenamefont {Sui}}]{Cheng2010}%
  \BibitemOpen
  \bibfield  {author} {\bibinfo {author} {\bibfnamefont {J.-G.}\ \bibnamefont
  {Cheng}}, \bibinfo {author} {\bibfnamefont {F.}~\bibnamefont {Zhou}},
  \bibinfo {author} {\bibfnamefont {J.-S.}\ \bibnamefont {Zhou}}, \bibinfo
  {author} {\bibfnamefont {J.~B.}\ \bibnamefont {Goodenough}},\ and\ \bibinfo
  {author} {\bibfnamefont {Y.}~\bibnamefont {Sui}},\ }\bibfield  {title}
  {\bibinfo {title} {{Enhanced thermoelectric power near the quantum phase
  transition in the itinerant-electron ferromagnet MnSi}},\ }\href
  {https://doi.org/10.1103/PhysRevB.82.214402} {\bibfield  {journal} {\bibinfo
  {journal} {Phys. Rev. B}\ }\textbf {\bibinfo {volume} {82}},\ \bibinfo
  {pages} {214402} (\bibinfo {year} {2010})}\BibitemShut {NoStop}%
\bibitem [{\citenamefont {Hirokane}\ \emph {et~al.}(2016)\citenamefont
  {Hirokane}, \citenamefont {Tomioka}, \citenamefont {Imai}, \citenamefont
  {Maeda},\ and\ \citenamefont {Onose}}]{Hirokane2016}%
  \BibitemOpen
  \bibfield  {author} {\bibinfo {author} {\bibfnamefont {Y.}~\bibnamefont
  {Hirokane}}, \bibinfo {author} {\bibfnamefont {Y.}~\bibnamefont {Tomioka}},
  \bibinfo {author} {\bibfnamefont {Y.}~\bibnamefont {Imai}}, \bibinfo {author}
  {\bibfnamefont {A.}~\bibnamefont {Maeda}},\ and\ \bibinfo {author}
  {\bibfnamefont {Y.}~\bibnamefont {Onose}},\ }\bibfield  {title} {\bibinfo
  {title} {{Longitudinal and transverse thermoelectric transport in MnSi}},\
  }\href {https://doi.org/10.1103/PhysRevB.93.014436} {\bibfield  {journal}
  {\bibinfo  {journal} {Phys. Rev. B}\ }\textbf {\bibinfo {volume} {93}},\
  \bibinfo {pages} {014436} (\bibinfo {year} {2016})}\BibitemShut {NoStop}%
\bibitem [{\citenamefont {Zhang}\ and\ \citenamefont {Li}(2004)}]{Zhang2004}%
  \BibitemOpen
  \bibfield  {author} {\bibinfo {author} {\bibfnamefont {S.}~\bibnamefont
  {Zhang}}\ and\ \bibinfo {author} {\bibfnamefont {Z.}~\bibnamefont {Li}},\
  }\bibfield  {title} {\bibinfo {title} {{Roles of Nonequilibrium Conduction
  Electrons on the Magnetization Dynamics of Ferromagnets}},\ }\href
  {https://doi.org/10.1103/PhysRevLett.93.127204} {\bibfield  {journal}
  {\bibinfo  {journal} {Phys. Rev. Lett.}\ }\textbf {\bibinfo {volume} {93}},\
  \bibinfo {pages} {127204} (\bibinfo {year} {2004})}\BibitemShut {NoStop}%
\bibitem [{\citenamefont {Prychynenko}\ \emph {et~al.}(2018)\citenamefont
  {Prychynenko}, \citenamefont {Sitte}, \citenamefont {Litzius}, \citenamefont
  {Kr{\"u}ger}, \citenamefont {Bourianoff}, \citenamefont {Kl{\"a}ui},
  \citenamefont {Sinova},\ and\ \citenamefont
  {Everschor-Sitte}}]{Prychynenko2018}%
  \BibitemOpen
  \bibfield  {author} {\bibinfo {author} {\bibfnamefont {D.}~\bibnamefont
  {Prychynenko}}, \bibinfo {author} {\bibfnamefont {M.}~\bibnamefont {Sitte}},
  \bibinfo {author} {\bibfnamefont {K.}~\bibnamefont {Litzius}}, \bibinfo
  {author} {\bibfnamefont {B.}~\bibnamefont {Kr{\"u}ger}}, \bibinfo {author}
  {\bibfnamefont {G.}~\bibnamefont {Bourianoff}}, \bibinfo {author}
  {\bibfnamefont {M.}~\bibnamefont {Kl{\"a}ui}}, \bibinfo {author}
  {\bibfnamefont {J.}~\bibnamefont {Sinova}},\ and\ \bibinfo {author}
  {\bibfnamefont {K.}~\bibnamefont {Everschor-Sitte}},\ }\bibfield  {title}
  {\bibinfo {title} {{Magnetic skyrmion as a nonlinear resistive element: a
  potential building block for reservoir computing}},\ }\href
  {https://doi.org/10.1103/PhysRevApplied.9.014034} {\bibfield  {journal}
  {\bibinfo  {journal} {Phys. Rev. Appl.}\ }\textbf {\bibinfo {volume} {9}},\
  \bibinfo {pages} {014034} (\bibinfo {year} {2018})}\BibitemShut {NoStop}%
\end{thebibliography}%

\end{document}